\documentclass[12pt]{article}
\usepackage{amsfonts,graphicx,amsmath,amsthm,amssymb,epsfig}
\usepackage{float}
\allowdisplaybreaks
\begin{document}

\title{\bf Anisotropic Stellar Models with Tolman IV Spacetime in Non-minimally Coupled Theory}
\author{M. Sharif$^1$ \thanks{msharif.math@pu.edu.pk}~ and Tayyab Naseer$^{1,2}$
\thanks{tayyabnaseer48@yahoo.com; tayyab.naseer@math.uol.edu.pk}\\
$^1$ Department of Mathematics and Statistics, The University of Lahore,\\
1-KM Defence Road Lahore, Pakistan.\\
$^2$ Department of Mathematics, University of the Punjab,\\
Quaid-e-Azam Campus, Lahore-54590, Pakistan.}

\date{}
\maketitle

\begin{abstract}
This article aims to investigate various anisotropic stellar models
in the background of $f(\mathcal{R},\mathcal{T},\mathcal{Q})$
gravity, where
$\mathcal{Q}=\mathcal{R}_{\varphi\vartheta}\mathcal{T}^{\varphi\vartheta}$.
In this regard, we adopt two standard models as
$\mathcal{R}+\zeta\mathcal{Q}$ and
$\mathcal{R}+\zeta\mathcal{R}\mathcal{Q}$, where $\zeta$ symbolizes
an arbitrary coupling parameter. We take spherical interior geometry
and find solution to the modified gravitational field equations
corresponding to each model by employing the `Tolman IV' spacetime.
We need an additional constraint to close the system of field
equations, thus the $\mathbb{MIT}$ bag model equation of state is
chosen. The effects of modified theory on physical properties of six
compact stars like PSR J 1614 2230,~SMC X-1,~Cen X-3,~PSR J
1903+327,~SAX J 1808.4-3658 and 4U 1820-30 are analyzed by using
their respective masses and radii. We also determine the values of
three unknowns involving in Tolman IV solution as well as the bag
constant for each star at the hypersurface. Furthermore, various
characteristics of the resulting solutions are examined through
graphical interpretation for $\zeta=\pm5$. Finally, we explore the
stability of the compact objects through two different approaches.
We conclude that our model-I produces physically acceptable
structures corresponding to each star candidate for both values of
$\zeta$ whereas model-II is stable only for $\zeta=5$.
\end{abstract}
{\bf Keywords:}
$f(\mathcal{R},\mathcal{T},\mathcal{R}_{\varphi\vartheta}\mathcal{T}^{\varphi\vartheta})$
gravity; Stellar models; Anisotropy. \\
{\bf PACS:} 04.20.Jb; 98.80.Jk; 03.50.De.

\section{Introduction}

General theory of relativity ($\mathbb{GR}$) is observed as the most
acceptable gravitational theory in the scientific community till
date to address various challenges. However, it is unsatisfactorily
enough to get better understanding of the cosmic' rapid expansion.
Scientists recently postulated various modifications of
$\mathbb{GR}$ to get a theory that can tackle the perplexing issues
associated with a cosmic evolution like its rapid expansion as well
as the dark matter etc. Different recent observations indicate the
existence of a force possessing anti-gravity effects that causes
such an accelerated expansion. This force is termed as the dark
energy due to which galactic structures move apart from each other.
The $f(\mathcal{R})$ theory is considered as the straightforward
extension of $\mathbb{GR}$ that was established by modifying the
Einstein-Hilbert action in which the Ricci scalar $\mathcal{R}$ is
replaced by its generic function \cite{1}. Various techniques have
been employed in this context to explore the stability and viability
of celestial structures \cite{2}-\cite{2f}. Many cosmological
concerns like late-time era of the universe' evolution \cite{3}, the
inflationary epoch \cite{4} and the background of cosmic development
\cite{5,5a} have been discussed by using different $f(\mathcal{R})$
gravity models.

Earlier, Bertolami et al. \cite{10} were interested in discovering
more fascinating features of the cosmos. They included the effects
of geometric entities into the matter Lagrangian in $f(\mathcal{R})$
theory so that the notion of the matter-geometry interaction can be
introduced. Various astronomers have been inspired from such
couplings that make them able to put their attention on the study of
rapid universe's expansion. This idea was recently extended at the
action level to develop some modified theories of gravity that
eventually becomes a very important topic for astrophysicists. Harko
et al. \cite{20} suggested the first theory based on this concept,
called $f(\mathcal{R},\mathcal{T})$ gravity. The entity
$\mathcal{T}$ in the functional indicates trace of the
energy-momentum tensor $(\mathbb{EMT})$ whose incorporation leads to
the non-conservation phenomenon. The physical feasibility of massive
self-gravitating structures have been investigated in this framework
and several remarkable results were analyzed \cite{21}-\cite{21h}.
When an inner configuration of a star incorporates matter for which
$\mathcal{T}=0$, this theory cannot be able to preserve the coupling
effects.

A more complex functional depending on $\mathcal{R},~\mathcal{T}$
and $\mathcal{R}_{\phi\psi}\mathcal{T}^{\phi\psi}$ was established
by Haghani et al. \cite{22} to remove the limitations of
$f(\mathcal{R},\mathcal{T})$ theory. A cosmic era, in which our
universe expanded exponentially, has been studied in this framework.
They studied three different
$f(\mathcal{R},\mathcal{T},\mathcal{Q})$ models to analyze their
physical feasibility and calculated conserved $\mathbb{EMT}$ even in
this theory by employing Lagrange multiplier technique. Two
different choices like $\mathbb{L}_m=\mu$ and $-P$ have been adopted
by Sharif and Zubair \cite{22a} to discuss different properties of
black holes along with their thermodynamical laws corresponding to
two mathematical models such as $\mathcal{R}+\zeta\mathcal{Q}$ and
$\mathcal{R}(1+\zeta\mathcal{Q})$, and determined viability
constraints for these models.

Odintsov and S\'{a}ez-G\'{o}mez \cite{23} studied different
cosmological models in $f(\mathcal{R},\mathcal{T},\mathcal{Q})$
theory and found that the non-constant matter distributions may
produce pure de Sitter universe in this context. By adopting
suitable scalar/vector fields, Ayuso et al. \cite{24} revealed that
modified field equations are highly non-linear due to the occurrence
of conformal and arbitrary non-minimal matter-geometry couplings.
The numerical solutions of different models and their stability
criteria has also been discussed by choosing a perturbation function
\cite{25}. Sharif and Waseem \cite{25a} have chosen matter
Lagrangian in terms of different pressure components to calculate
anisotropic solutions and concluded that different choices of
$\mathbb{L}_m$ yield significantly different results. The curvature
tensor has been orthogonally decomposed in relation to
$f(\mathcal{R},\mathcal{T},\mathcal{Q})$ $\mathbb{EMT}$ and its
trace to get some scalar functions for charged/uncharged fluids
which play an important role in discussing massive systems
\cite{26}-\cite{26e}. We have obtained several anisotropic solutions
to the field equations by utilizing different techniques and
observed them acceptable \cite{27a}-\cite{27d}. Some other modified
theories such as $\kappa(\mathcal{R},\mathcal{T})$,
$f(Q,\mathcal{T})$ with $Q$ being the non-metricity scalar, etc.
have also been discussed in the literature \cite{27e}-\cite{27g}.

Among the myriad inscrutable cosmic' constituents, stars are
recognized as elementary components of our galaxy. Many
astrophysicists were motivated by the composition of such structures
which led them to fully concentrate on the evolutionary stages of
compact stars. The intriguing features of neutron stars attracted
much amongst all compact bodies resulting from a gravitational
collapse. A neutron star has a mass roughly 1 to 3 times that of a
solar system ($M_{\bigodot}$) and contains newly created neutrons in
its core. These neutrons allow to fork out a pressure to offset the
attraction of gravity and thus further collapse is prevented. A
neutron star was firstly predicted in 1934 \cite{28}, however it was
observationally verified after a couple of years because these
structures are often indiscernible and do not emit sufficient
radiations. The composition of another hypothetical structure
(between black hole and a neutron star) is known as quark star which
contains strange, down and up quark matter configurations in its
interior \cite{29a}-\cite{29h}.

The state determinants representing interior of self-gravitating
(isotropic and anisotropic) bodies usually contain energy density
and pressure. These attributes can be interconnected by means of
some particular constraints. A well-known constraint in this regard
is the $\mathbb{MIT}$ bag model equation of state
($\mathbb{E}$o$\mathbb{S}$) \cite{30}. It has also been pointed out
that such an equation efficiently describes the compactness of
strange systems like 4U 1820-30, PSR 0943+10, RXJ 185635-3754, Her
X-1, 4U 1728-34 and SAX J 1808.4-3658, etc., however, this cannot be
done through an $\mathbb{E}$o$\mathbb{S}$ for neutron stars
\cite{33a}. Several physicists employed this model in order to
analyze the physical composition of quark star like bodies
\cite{33b}-\cite{34aa}. Demorest et al. \cite{34b} discussed
acompact strange structure PSR J1614-2230 and observed that
$\mathbb{MIT}$ model is the only equation that supports such heavily
objects. Rahaman et al. \cite{35} studied the composition of some
particular stars by using this model with an interpolating function.
Sharif and his associates \cite{38,38c} extended this work in
various modified scenarios and developed stable models.

Several approaches have been proposed in the literature that help in
formulating solutions to the Einstein/modified field equations. For
instance, a particular $\mathbb{E}$o$\mathbb{S}$ \cite{38h,38i}, the
known form of the metric potentials \cite{38k} or a complexity-free
condition \cite{38l} etc. can be used in this context. Among several
metric potentials proposed in the literature, Tolman IV ansatz has
prompted many researchers to use this in studying anisotropic
compact structures. Ovalle and Linares \cite{38m} developed this
solution in the braneworld scenario and provided an evidence in the
favor of the reduction in the compactness of stellar body due to the
impact of bulk. Singh et al. \cite{38n} formulated charged Tolman IV
anisotropic fluid solutions and found them advantageous to model
neutron as well as quark stars. Some other charged/uncharged works
related to this gravitational potential has been done in
$\mathbb{GR}$ and modified theories through the gravitational
decoupling, complexity factor, etc. \cite{38o}-\cite{38r}.

This paper aims to analyze the physical acceptability of the
developed Tolman IV anisotropic solutions to
$f(\mathcal{R},\mathcal{T},\mathcal{R}_{\varphi\vartheta}\mathcal{T}^{\varphi\vartheta})$
field equations. The paper is structured as follows. Some
fundamentals of this modified theory and the corresponding field
equations for two different models \big(such as $\mathcal{R}+\zeta
\mathcal{R}_{\varphi\vartheta}\mathcal{T}^{\varphi\vartheta}$ and
$\mathcal{R}(1+\zeta
\mathcal{R}_{\varphi\vartheta}\mathcal{T}^{\varphi\vartheta})$\big)
are presented in section \textbf{2}. We also consider $\mathbb{MIT}$
bag model $\mathbb{E}$o$\mathbb{S}$ to characterize the quarks'
interior. Section \textbf{3} considers the Tolman IV spacetime to
make the metric potentials known. Further, the unknown triplet
$(A,B,C)$ in this solution are determined through junction
conditions. The physical attributes of the resulting solutions for
different quark stars are graphically checked in section \textbf{4}
for a fixed value of the model parameter. Lastly, section \textbf{5}
summarizes all our results.

\section{The $f(\mathcal{R},\mathcal{T},\mathcal{R}_{\varphi\vartheta}\mathcal{T}^{\varphi\vartheta})$ Theory}

The action of this modified theory (with $\kappa=8\pi$) has the form
\cite{23}
\begin{equation}\label{g1}
\mathbb{S}_{f(\mathcal{R},\mathcal{T},\mathcal{R}_{\varphi\vartheta}\mathcal{T}^{\varphi\vartheta})}=\int\sqrt{-g}
\left\{\frac{f(\mathcal{R},\mathcal{T},\mathcal{R}_{\varphi\vartheta}\mathcal{T}^{\varphi\vartheta})}{16\pi}
+\mathbb{L}_{m}\right\}d^{4}x,
\end{equation}
which involves $\mathbb{L}_{m}$ representing matter Lagrangian of
the fluid distribution and we take it as $\mathbb{L}_{m}=-\mu$,
$\mu$ is the energy density. The field equations can be obtained
from the action \eqref{g1} through the variational principle as
\begin{equation}\label{g2}
\mathcal{G}_{\varphi\vartheta}=8\pi\mathcal{T}_{\varphi\vartheta}^{(\mathrm{EFF})}=\frac{8\pi\mathcal{T}_{\varphi\vartheta}}
{f_{\mathcal{R}}-\mathbb{L}_{m}f_{\mathcal{Q}}}+\mathcal{T}_{\varphi\vartheta}^{(\mathcal{C})},
\end{equation}
which couples the spacetime geometry with matter distribution. The
terms $\mathcal{G}_{\varphi\vartheta}$ and
$\mathcal{T}_{\varphi\vartheta}^{(\mathrm{EFF})}$ in the above
equation are known as the Einstein tensor and the effective
$\mathbb{EMT}$ in this theory. The sector
$\mathcal{T}_{\varphi\vartheta}^{(\mathcal{C})}$ represents the
correction terms of extended gravity that becomes
\begin{eqnarray}\nonumber
\mathcal{T}_{\varphi\vartheta}^{(\mathcal{C})}&=&-\frac{1}{\bigg(\mathbb{L}_{m}f_{\mathcal{Q}}-f_{\mathcal{R}}\bigg)}
\left[\left(f_{\mathcal{T}}+\frac{1}{2}\mathcal{R}f_{\mathcal{Q}}\right)\mathcal{T}_{\varphi\vartheta}
-\left\{\mathbb{L}_{m}f_{\mathcal{T}}-\frac{\mathcal{R}}{2}\bigg(\frac{f}{\mathcal{R}}-f_{\mathcal{R}}\bigg)\right.\right.\\\nonumber
&+&\left.\frac{1}{2}\nabla_{\varrho}\nabla_{\omega}(f_{\mathcal{Q}}\mathcal{T}^{\varrho\omega})\right\}g_{\varphi\vartheta}
-\frac{1}{2}\Box(f_{\mathcal{Q}}\mathcal{T}_{\varphi\vartheta})-(g_{\varphi\vartheta}\Box-
\nabla_{\varphi}\nabla_{\vartheta})f_{\mathcal{R}}\\\label{g4}
&-&2f_{\mathcal{Q}}\mathcal{R}_{\varrho(\varphi}\mathcal{T}_{\vartheta)}^{\varrho}
+\nabla_{\varrho}\nabla_{(\varphi}[\mathcal{T}_{\vartheta)}^{\varrho}
f_{\mathcal{Q}}]+2(f_{\mathcal{Q}}\mathcal{R}^{\varrho\omega}+\left.f_{\mathcal{T}}g^{\varrho\omega})\frac{\partial^2\mathbb{L}_{m}}
{\partial g^{\varphi\vartheta}\partial g^{\varrho\omega}}\right],
\end{eqnarray}
where $f_{\mathcal{R}}=\frac{\partial
f(\mathcal{R},\mathcal{T},\mathcal{Q})}{\partial
\mathcal{R}},~f_{\mathcal{T}}=\frac{\partial
f(\mathcal{R},\mathcal{T},\mathcal{Q})}{\partial \mathcal{T}}$ and
$f_{\mathcal{Q}}=\frac{\partial
f(\mathcal{R},\mathcal{T},\mathcal{Q})}{\partial \mathcal{Q}}$.
Also, $\nabla_\varrho$ and $\Box\equiv
\frac{1}{\sqrt{-g}}\partial_\varphi\big(\sqrt{-g}g^{\varphi\vartheta}\partial_{\vartheta}\big)$
represent covariant derivative and the D'Alambert operator,
respectively. The term engaging matter Lagrangian in Eq.\eqref{g4}
leads to $\frac{\partial^2\mathbb{L}_{m}} {\partial
g^{\varphi\vartheta}\partial g^{\varrho\omega}}=0$ \cite{22}. The
involvement of the $\mathbb{EMT}$ in the functional
$f(\mathcal{R},\mathcal{T},\mathcal{Q})$ results in its
non-conserved nature (i.e., $\nabla_\varphi
\mathcal{T}^{\varphi\vartheta}\neq 0$) and thus this theory violates
the equivalence principle opposite to other extended theories
\cite{39,40}. Consequently, an extra force is exerted on particles
that triggers their non-geodesic motion in the gravitational field
of a celestial object, and hence we obtain
\begin{align}\nonumber
\nabla^\varphi
\mathcal{T}_{\varphi\vartheta}&=\frac{2}{2f_\mathcal{T}+\mathcal{R}f_\mathcal{Q}+16\pi}\bigg[\nabla_\varphi
\big(f_\mathcal{Q}\mathcal{R}^{\varrho\varphi}\mathcal{T}_{\varrho\vartheta}\big)-\mathcal{G}_{\varphi\vartheta}\nabla^\varphi
\big(f_\mathcal{Q}\mathbb{L}_m\big)+\nabla_\vartheta
\big(\mathbb{L}_mf_\mathcal{T}\big)\\\label{g4a}
&-\frac{1}{2}\nabla_\vartheta\mathcal{T}^{\varrho\omega}\big(f_\mathcal{T}g_{\varrho\omega}+f_\mathcal{Q}\mathcal{R}_{\varrho\omega}\big)
-\frac{1}{2}\big\{\nabla^{\varphi}(\mathcal{R}f_{\mathcal{Q}})+2\nabla^{\varphi}f_{\mathcal{T}}\big\}\mathcal{T}_{\varphi\vartheta}\bigg].
\end{align}

The $\mathbb{EMT}$ is primarily used to study the internal
composition and nature of fluid distribution of a self-gravitating
system. A large number of celestial objects among the broad range of
massive structures in our cosmos are thought to be interconnected
with pressure anisotropy, therefore this element becomes significant
for astrophysicists in studying the evolution of stellar models. In
this context, the anisotropic $\mathbb{EMT}$ is given as
\begin{equation}\label{g5}
\mathcal{T}_{\varphi\vartheta}=(\mu+P_\bot) \mathcal{K}_{\varphi}
\mathcal{K}_{\vartheta}+P_\bot
g_{\varphi\vartheta}+\left(P_r-P_\bot\right)\mathcal{W}_\varphi\mathcal{W}_\vartheta,
\end{equation}
where the state variables $\mu,~P_\bot$ and $P_r$ stand for energy
density, tangential pressure and radial pressure, respectively.
Moreover, $\mathcal{W}_{\varphi}$ is the four-vector and
$\mathcal{K}_\varphi$ indicates the four-velocity. The
$f(\mathcal{R},\mathcal{T},\mathcal{Q})$ field equations provide
their trace as
\begin{align}\nonumber
&3\nabla^{\varrho}\nabla_{\varrho}
f_\mathcal{R}-\mathcal{T}(8\pi+f_\mathcal{T})-\mathcal{R}\left(\frac{\mathcal{T}}{2}f_\mathcal{Q}-f_\mathcal{R}\right)+\frac{1}{2}
\nabla^{\varrho}\nabla_{\varrho}(f_\mathcal{Q}\mathcal{T})\\\nonumber
&+\nabla_\varphi\nabla_\varrho(f_\mathcal{Q}\mathcal{T}^{\varphi\varrho})
-2f+(\mathcal{R}f_\mathcal{Q}+4f_\mathcal{T})\mathbb{L}_m+2\mathcal{R}_{\varphi\varrho}\mathcal{T}^{\varphi\varrho}f_\mathcal{Q}\\\nonumber
&-2g^{\vartheta\xi} \frac{\partial^2\mathbb{L}_m}{\partial
g^{\vartheta\xi}\partial
g^{\varphi\varrho}}\left(f_\mathcal{T}g^{\varphi\varrho}+f_\mathcal{Q}R^{\varphi\varrho}\right)=0.
\end{align}
The $f(\mathcal{R},\mathcal{T})$ gravity can be achieved from the
overhead equation by putting $f_\mathcal{Q}=0$, while
$f_\mathcal{T}=0$ yields $f(\mathcal{R})$ theory.

We take interior geometry represented by static spherically
symmetric metric as
\begin{equation}\label{g6}
ds^2=-e^{\lambda} dt^2+e^{\xi} dr^2+r^2d\theta^2+r^2\sin^2\theta
d\phi^2,
\end{equation}
where $\lambda=\lambda(r)$ and $\xi=\xi(r)$. The four-vector and
four-velocity become
\begin{equation}\label{g7}
\mathcal{W}^\varphi=\delta^\varphi_1 e^{\frac{-\xi}{2}}, \quad
\mathcal{K}^\varphi=\delta^\varphi_0 e^{\frac{-\lambda}{2}},
\end{equation}
which must satisfy $\mathcal{K}^\varphi \mathcal{K}_{\varphi}=-1$
and $\mathcal{W}^\varphi \mathcal{K}_{\varphi}=0$. Our cosmos is
currently experiencing rapidly expansion phase and comprising of a
large number of stars existing in non-linear reign, yet analyzing
their linear behavior can help astronomers to understand formation
of these massive structures in a proper way. On the other hand, the
inclusion of
$\mathcal{R}_{\varphi\vartheta}\mathcal{T}^{\varphi\vartheta}$ in
$f(\mathcal{R},\mathcal{T},\mathcal{Q})$ gravity makes it enormously
much complicated in contrast with $f(\mathcal{R},\mathbb{L}_m)$ and
$f(\mathcal{R},\mathcal{T})$ theories. In this regard, we consider
the following two concrete functional forms, namely \cite{22}
\begin{itemize}
\item Model-I:\quad $f(\mathcal{R},\mathcal{T},\mathcal{R}_{\varphi\vartheta}\mathcal{T}^{\varphi\vartheta})=\mathcal{R}+\zeta
\mathcal{R}_{\varphi\vartheta}\mathcal{T}^{\varphi\vartheta}$,
\item Model-II:\quad $f(\mathcal{R},\mathcal{T},\mathcal{R}_{\varphi\vartheta}\mathcal{T}^{\varphi\vartheta})=\mathcal{R}(1+\zeta
\mathcal{R}_{\varphi\vartheta}\mathcal{T}^{\varphi\vartheta})$,
\end{itemize}
where $\zeta$ indicates a coupling parameter. It is remarkable to
note that different values of the coupling parameter (found to be
within the observed range) guarantee the physical feasibility of the
respective models. Haghani et al. \cite{22} discussed these models
to study evolution of the scale factor as well as deceleration
parameter. The isotropic distributions have been discussed by
considering model-I from which certain acceptable values of the
coupling parameter are deduced \cite{22a}. The quantity
$\mathcal{Q}=\mathcal{R}_{\varphi\vartheta}\mathcal{T}^{\varphi\vartheta}$
in the above models becomes
\begin{eqnarray}\nonumber
\mathcal{Q}&=&e^{-\xi}\bigg[\frac{\mu}{4}\left(\lambda'^2-\lambda'\xi'+2\lambda''+\frac{4\lambda'}{r}\right)
-\frac{P_r}{4}\left(\lambda'^2-\lambda'\xi'+2\lambda''+\frac{4\xi'}{r}\right)\\\nonumber
&+&P_\bot
\left(\frac{\xi'}{r}-\frac{\lambda'}{r}+\frac{2e^\xi}{r^2}-\frac{2}{r^2}\right)\bigg].
\end{eqnarray}
Here, $'=\frac{\partial}{\partial r}$.

The field equations \eqref{g2} for fluid \eqref{g5} corresponding to
the model-I are given as
\begin{align}\nonumber
8\pi\mu&=e^{-\xi}\bigg[\frac{\xi'}{r}+\frac{e^\xi}{r^2}-\frac{1}{r^2}+\zeta\bigg\{\mu\bigg(\frac{3\lambda'\xi'}{8}-\frac{\lambda'^2}{8}
+\frac{\xi'}{r}+\frac{e^\xi}{r^2}-\frac{3\lambda''}{4}-\frac{3\lambda'}{2r}\\\nonumber
&-\frac{1}{r^2}\bigg)-\mu'\bigg(\frac{\xi'}{4}-\frac{1}{r}-\lambda'\bigg)+\frac{\mu''}{2}+P_r\bigg(\frac{\lambda'\xi'}{8}
-\frac{\lambda'^2}{8}-\frac{\lambda''}{4}+\frac{\xi'}{2r}+\frac{\xi''}{2}\\\label{g8}
&-\frac{3\xi'^2}{4}\bigg)+\frac{5\xi'P'_r}{4}-\frac{P''_r}{2}+P_\bot\bigg(\frac{\xi'}{2r}-\frac{\lambda'}{2r}+\frac{3e^\xi}{r^2}
-\frac{1}{r^2}\bigg)-\frac{P'_\bot}{r}\bigg\}\bigg],\\\nonumber 8\pi
P_r&=e^{-\xi}\bigg[\frac{\lambda'}{r}-\frac{e^\xi}{r^2}+\frac{1}{r^2}+\zeta\bigg\{\mu\bigg(\frac{\lambda'\xi'}{8}+\frac{\lambda'^2}{8}
-\frac{\lambda''}{4}-\frac{\lambda'}{2r}\bigg)-\frac{\lambda'\mu'}{4}-P_r\\\nonumber
&\times\bigg(\frac{5\lambda'^2}{8}-\frac{7\lambda'\xi'}{8}+\frac{5\lambda''}{4}-\frac{7\xi'}{2r}+\frac{\lambda'}{r}-\xi'^2
-\frac{e^\xi}{r^2}+\frac{1}{r^2}\bigg)+P'_r\bigg(\frac{\lambda'}{4}+\frac{1}{r}\bigg)\\\label{g8a}
&-P_\bot\bigg(\frac{\xi'}{2r}-\frac{\lambda'}{2r}+\frac{3e^\xi}{r^2}-\frac{1}{r^2}\bigg)+\frac{P'_\bot}{r}\bigg\}\bigg],\\\nonumber
8\pi
P_\bot&=e^{-\xi}\bigg[\frac{\lambda'^2}{4}-\frac{\lambda'\xi'}{4}+\frac{\lambda''}{2}-\frac{\xi'}{2r}+\frac{\lambda'}{2r}
+\zeta\bigg\{\mu\bigg(\frac{\lambda'^2}{8}+\frac{\lambda'\xi'}{8}-\frac{\lambda''}{4}-\frac{\lambda'}{2r}\bigg)\\\nonumber
&-\frac{\lambda'\mu'}{4}+P_r\bigg(\frac{\lambda'^2}{8}-\frac{\lambda'\xi'}{8}+\frac{\lambda''}{4}-\frac{\xi'}{2r}-\frac{\xi''}{2}
+\frac{3\xi'^2}{4}\bigg)-\frac{5\xi'P'_r}{4}+\frac{P''_r}{2}\\\label{g8b}
&-P_\bot\bigg(\frac{\lambda'^2}{4}-\frac{\lambda'\xi'}{4}+\frac{\lambda''}{2}-\frac{\xi'}{r}+\frac{\lambda'}{r}\bigg)
-P'_\bot\bigg(\frac{\xi'}{4}-\frac{\lambda'}{4}-\frac{3}{r}\bigg)+\frac{P''_\bot}{2}\bigg\}\bigg],
\end{align}
and for the model-II, we have
\begin{align}\nonumber
8\pi\mu&=e^{-\xi}\bigg[\frac{\xi'}{r}+\frac{e^\xi}{r^2}-\frac{1}{r^2}+\zeta\bigg\{\mu\bigg(\bigg(\frac{\xi'}{r}+\frac{e^\xi}{r^2}
-\frac{1}{r^2}\bigg)\beta_1-\mathcal{R}\bigg(\frac{1}{r^2}-\frac{\xi'}{r}-\frac{e^\xi}{r^2}\\\nonumber
&+\frac{\mathcal{R}e^\xi}{2}-\frac{3\lambda'^2}{8}-\frac{3\lambda'}{2r}+\frac{5\lambda'\xi'}{8}-\frac{3\lambda''}{4}\bigg)
+\mathcal{R}'\bigg(\frac{\xi'}{2}-\frac{1}{r}\bigg)-\frac{\mathcal{R}''}{2}-\beta_4\bigg(\frac{2}{r}\\\nonumber
&-\frac{\xi'}{2}\bigg)-\beta_7\bigg)+\mu'\bigg(\beta_1\bigg(\frac{\xi'}{2}-\frac{2}{r}\bigg)-\mathcal{R}\bigg(\frac{1}{r}-\frac{\xi'}{4}\bigg)
-\mathcal{R}'-2\beta_4\bigg)-\mu''\bigg(\beta_1\\\nonumber
&+\frac{\mathcal{R}}{2}\bigg)-P_r\bigg(\bigg(\frac{1}{r^2}-\frac{\xi'}{r}-\frac{e^\xi}{r^2}\bigg)\beta_2+\mathcal{R}\bigg(\frac{\lambda'^2}{8}
-\frac{1}{r^2}-\frac{\lambda'\xi'}{8}+\frac{\xi'}{2r}+\frac{\lambda''}{4}\bigg)\\\nonumber
&-\mathcal{R}'\bigg(\frac{2}{r}-\frac{\xi'}{2}\bigg)-\frac{\mathcal{R}''}{2}+\beta_5\bigg(\frac{2}{r}-\frac{\xi'}{2}\bigg)+\beta_8\bigg)
-P'_r\bigg(\beta_2\bigg(\frac{2}{r}-\frac{\xi'}{2}\bigg)+\mathcal{R}\\\nonumber
&\times\bigg(\frac{\xi'}{4}-\frac{2}{r}\bigg)-\mathcal{R}'+2\beta_5\bigg)-P''_r\bigg(\beta_2-\frac{\mathcal{R}}{2}\bigg)
-P_\bot\bigg(\bigg(\frac{1}{r^2}-\frac{\xi'}{r}-\frac{e^\xi}{r^2}\bigg)\beta_3\\\nonumber
&+\mathcal{R}\bigg(\frac{\lambda'}{2r}+\frac{1}{r^2}-\frac{\xi'}{2r}\bigg)+\frac{\mathcal{R}'}{r}+\beta_6\bigg(\frac{2}{r}-\frac{\xi'}{2}\bigg)
+\beta_9\bigg)-P'_\bot\bigg(\beta_3\bigg(\frac{2}{r}-\frac{\xi'}{2}\bigg)\\\label{g8c}
&+\frac{\mathcal{R}}{2}+2\beta_6\bigg)-P''_\bot\beta_3\bigg\}\bigg],\\\nonumber
8\pi
P_r&=e^{-\xi}\bigg[\frac{\lambda'}{r}-\frac{e^\xi}{r^2}+\frac{1}{r^2}+\zeta\bigg\{\mu\bigg(\bigg(\frac{\lambda'}{r}-\frac{e^\xi}{r^2}
+\frac{1}{r^2}\bigg)\beta_1-\mathcal{R}\bigg(-\frac{1}{r^2}-\frac{\lambda'}{r}+\frac{e^\xi}{r^2}\\\nonumber
&+\frac{\lambda'^2}{8}+\frac{\lambda'}{2r}-\frac{\lambda'\xi'}{8}+\frac{\lambda''}{4}\bigg)
+\frac{\mathcal{R}'\lambda'}{4}+\beta_4\bigg(\frac{2}{r}+\frac{\lambda'}{2}\bigg)\bigg)+\mu'\bigg(\beta_1\bigg(\frac{2}{r}
+\frac{\lambda'}{2}\bigg)\\\nonumber
&+\frac{\mathcal{R}\lambda'}{4}\bigg)+P_r\bigg(\bigg(\frac{1}{r^2}+\frac{\lambda'}{r}-\frac{e^\xi}{r^2}\bigg)\beta_2
-\mathcal{R}\bigg(\frac{\mathcal{R}e^\xi}{2}-\frac{3\lambda'^2}{8}+\frac{1}{r^2}+\frac{\lambda'}{r}+\frac{3\xi'}{2r}\\\nonumber
&+\frac{3\lambda'\xi'}{8}-\frac{3\lambda''}{4}\bigg)-\mathcal{R}'\bigg(\frac{1}{r}+\frac{\lambda'}{4}\bigg)
+\beta_5\bigg(\frac{2}{r}+\frac{\lambda'}{2}\bigg)\bigg)+P'_r\bigg(\beta_2\bigg(\frac{2}{r}+\frac{\lambda'}{2}\bigg)\\\nonumber
&-\mathcal{R}\bigg(\frac{\lambda'}{4}+\frac{1}{r}\bigg)\bigg)-P_\bot\bigg(\bigg(\frac{\lambda'}{r}+\frac{e^\xi}{r^2}-\frac{1}{r^2}\bigg)\beta_3
+\mathcal{R}\bigg(\frac{\xi'}{2r}-\frac{1}{r^2}-\frac{\lambda'}{2r}\bigg)+\frac{\mathcal{R}'}{r}\\\label{g8d}
&-\beta_6\bigg(\frac{2}{r}+\frac{\lambda'}{2}\bigg)\bigg)+P'_\bot\bigg(\beta_3\bigg(\frac{2}{r}
+\frac{\lambda'}{2}\bigg)-\frac{\mathcal{R}}{2}\bigg)\bigg\}\bigg],\\\nonumber
8\pi
P_\bot&=e^{-\xi}\bigg[\frac{\lambda''}{2}-\frac{\lambda'\xi'}{4}+\frac{\lambda'^2}{4}+\frac{\lambda'}{2r}-\frac{\xi}{2r}
+\zeta\bigg\{\mu\bigg(\beta_1\bigg(\frac{\lambda''}{2}-\frac{\lambda'\xi'}{4}+\frac{\lambda'^2}{4}+\frac{\lambda'}{2r}\\\nonumber
&-\frac{\xi}{2r}\bigg)-\mathcal{R}\bigg(\frac{\lambda'\xi'}{8}-\frac{\lambda''}{4}-\frac{\lambda'^2}{8}+\frac{\xi'}{2r}\bigg)
-\frac{\mathcal{R}'\lambda'}{4}+\beta_7-\bigg(\frac{\xi'}{2}-\frac{1}{r}-\frac{\lambda'}{2}\bigg)\beta_4\bigg)\\\nonumber
&-\mu'\bigg(\beta_1\bigg(\frac{\xi'}{2}-\frac{1}{r}-\frac{\lambda'}{2}\bigg)+\frac{\mathcal{R}\lambda'}{4}-2\beta_4\bigg)
+\mu''\beta_1-P_r\bigg(\bigg(\frac{\xi'}{2r}-\frac{\lambda''}{2}-\frac{\lambda'^2}{4}\\\nonumber
&-\frac{\lambda'}{2r}+\frac{\lambda'\xi'}{4}\bigg)\beta_2+\mathcal{R}\bigg(\frac{\lambda'^2}{8}+\frac{\lambda'}{2r}-\frac{\lambda'\xi'}{8}
+\frac{\lambda''}{4}\bigg)+\mathcal{R}'\bigg(\frac{\lambda'}{2}+\frac{1}{r}-\frac{\xi'}{4}\bigg)\\\nonumber
&+\frac{\mathcal{R}''}{2}+\beta_5\bigg(\frac{\xi'}{2}-\frac{1}{r}-\frac{\lambda'}{2}\bigg)
-\beta_8\bigg)-P'_r\bigg(\beta_2\bigg(\frac{\xi'}{2}-\frac{1}{r}-\frac{\lambda'}{2}\bigg)+\mathcal{R}\bigg(\frac{\lambda'}{2}\\\nonumber
&+\frac{1}{r}-\frac{\xi'}{4}\bigg)+\mathcal{R}'-2\beta_5\bigg)+P''_r\bigg(\beta_2-\frac{\mathcal{R}}{2}\bigg)
+P_\bot\bigg(\beta_3\bigg(\frac{\lambda''}{2}-\frac{\lambda'\xi'}{4}+\frac{\lambda'^2}{4}\\\nonumber
&+\frac{\lambda'}{2r}-\frac{\xi}{2r}\bigg)-\mathcal{R}\bigg(\frac{\mathcal{R}e^\xi}{2}-\frac{2}{r^2}+\frac{\xi'}{r}-\frac{\lambda'}{r}
+\frac{2e^\xi}{r^2}\bigg)-\mathcal{R}'\bigg(\frac{\lambda'}{4}-\frac{\xi'}{4}\bigg)-\frac{\mathcal{R}''}{2}\\\nonumber
&-\beta_6\bigg(\frac{\xi'}{2}-\frac{1}{r}-\frac{\lambda'}{2}\bigg)+\beta_9\bigg)-P'_\bot\bigg(\beta_3\bigg(\frac{\xi'}{2}-\frac{1}{r}
-\frac{\lambda'}{2}\bigg)+\mathcal{R}\bigg(\frac{\lambda'}{4}-\frac{\xi'}{4}+\frac{2}{r}\bigg)\\\label{g8e}
&+\mathcal{R}'-2\beta_6\bigg)+P''_\bot\bigg(\beta_3-\frac{\mathcal{R}}{2}\bigg)\bigg\}\bigg],
\end{align}
where the expressions of $\beta_i^{'s}$ (where $i=1,2,3,...,9$) are
given in Appendix $\mathbf{A}$.

The system of differential equations \eqref{g8}-\eqref{g8b} and
\eqref{g8c}-\eqref{g8e} involve much complications as they encompass
higher order derivatives of state variables and metric potentials,
thus it is problematic to find their corresponding solutions. The
mass of a sphere suggested by Misner-Sharp \cite{41b} has the form
\begin{equation}\nonumber
m(r)=\frac{r}{2}\big(1-g^{\varphi\vartheta}r_{,\varphi}r_{,\vartheta}\big),
\end{equation}
which results in
\begin{equation}\label{g12a}
m(r)=\frac{r}{2}\big(1-e^{-\xi}\big).
\end{equation}

Inside any geometrical configuration, there are various physical
variables such as pressure and energy density which can be
interconnected by making use of certain relations, named as
equations of state. The final outcome from dying a massive star can
be white dwarf, neutron star or black hole, among them neutron stars
are thought to be the most fascinating formations in our universe.
Some of these structures can further be transformed to quark stars
(black holes) according to lower (higher) densities in their cores
\cite{33b,41c,41d}, respectively. Meanwhile, these objects have much
dense cores and thus produce strong gravitational field around them
regardless of their small size. We have highly non-linear indefinite
systems \eqref{g8}-\eqref{g8b} and \eqref{g8c}-\eqref{g8e} as they
involve five unknowns \big($\lambda,\xi,\mu,P_\bot,P_r$\big), thus
we must need certain constraints in order to attain required
solutions. In this context, we assume a well-known $\mathbb{MIT}$
bag model $\mathbb{E}o\mathbb{S}$ to discuss physical aspects of
various quark candidates \cite{30}. The quark pressure is
\begin{equation}\label{g13}
P_r=\sum_{\iota=u,d,s}P^\iota-\mathfrak{B_c},
\end{equation}
where $\mathfrak{B_c}$ specifies the bag constant. In addition, the
above equation contains three superscripts $u,~d$ and $s$ which
correspond to up, down and strange matter, respectively. The energy
density and pressure for each matter inside quarks are linked
through the relation $\mu^\iota=3P^\iota$. The total energy density
in this case is
\begin{equation}\label{g14}
\mu=\sum_{\iota=u,d,s}\mu^\iota+\mathfrak{B_c}.
\end{equation}
Equations \eqref{g13} and \eqref{g14} provide the following equation
characterizing strange fluid as
\begin{equation}\label{g14a}
P_r=\frac{1}{3}\left(\mu-4\mathfrak{B_c}\right).
\end{equation}

Various physical properties of different star models have been
investigated by calculating their corresponding values of
$\mathfrak{B_c}$ \cite{41f,41h}, which are observed to be in their
acceptable ranges. The field equations \eqref{g8}-\eqref{g8b}
together with the $\mathbb{E}o\mathbb{S}$ \eqref{g14a} supply the
solution as
\begin{align}\nonumber
\mu&=\bigg[8\pi
e^{\xi}+\zeta\bigg(\frac{9\lambda''}{8}-\frac{e^{\xi}}{r^2}+\frac{1}{r^2}-\frac{\xi''}{8}-\frac{5\lambda'\xi'}{8}-\frac{\xi'^2}{16}
-\frac{7\xi'}{2r}+\frac{3\lambda'^2}{16}+\frac{7\lambda'}{4r}\bigg)\bigg]^{-1}\\\nonumber
&\times\bigg[\frac{3}{4}\bigg(\frac{\xi'}{r}+\frac{\lambda'}{r}\bigg)+\mathfrak{B_c}\bigg\{8\pi
e^\xi-\zeta\bigg(\frac{4\xi'}{r}-\frac{3\lambda'^2}{4}-\frac{3\lambda''}{2}+\frac{\xi''}{2}+\frac{\xi'^2}{4}+\lambda'\xi'\\\label{g14b}
&-\frac{\lambda'}{r}+\frac{e^\xi}{r^2}-\frac{1}{r^2}\bigg)\bigg\}\bigg],\\\nonumber
P_r&=\bigg[8\pi
e^{\xi}+\zeta\bigg(\frac{9\lambda''}{8}-\frac{e^{\xi}}{r^2}+\frac{1}{r^2}-\frac{\xi''}{8}-\frac{5\lambda'\xi'}{8}-\frac{\xi'^2}{16}
-\frac{7\xi'}{2r}+\frac{3\lambda'^2}{16}+\frac{7\lambda'}{4r}\bigg)\bigg]^{-1}\\\label{g14c}
&\times\bigg[\frac{1}{4}\bigg(\frac{\xi'}{r}+\frac{\lambda'}{r}\bigg)-\mathfrak{B_c}\bigg\{8\pi
e^\xi-\zeta\bigg(\frac{\lambda'\xi'}{2}
+\frac{\xi'}{r}-\frac{2\lambda'}{r}+\frac{e^\xi}{r^2}-\lambda''-\frac{1}{r^2}\bigg)\bigg\}\bigg],
\end{align}
whereas Eqs.\eqref{g8c}-\eqref{g8e} yield
\begin{align}\nonumber
\mu&=\bigg[\zeta\bigg\{\frac{3}{4}\bigg(\frac{\xi'}{r}+\frac{\lambda'}{r}\bigg)\bigg(\beta_1+\frac{\beta_2}{3}\bigg)
+\frac{3}{8}\big(\xi'+\lambda'\big)\bigg(\beta_4+\frac{\beta_5}{3}\bigg)
+\mathcal{R}\bigg(\frac{5\lambda'}{4r}-\frac{\mathcal{R}e^\xi}{2}\\\nonumber
&-\frac{7\lambda'\xi'}{16}+\frac{\lambda'^2}{16}+\frac{\lambda''}{2}+\frac{\xi'}{4r}\bigg)+\mathcal{R}'\bigg(\frac{\lambda'}{8}-\frac{1}{2r}
+\frac{\xi'}{8}\bigg)-\frac{\mathcal{R''}}{4}-\frac{\beta_8}{4}\bigg\}-8\pi
e^{\xi}\bigg]^{-1}\\\nonumber
&\times\bigg[-\frac{3}{4}\bigg(\frac{\xi'}{r}+\frac{\lambda'}{r}\bigg)-8\pi\mathfrak{B_c}e^\xi-\zeta\mathfrak{B_c}\bigg\{\bigg(\frac{\xi'}{r}
+\frac{\lambda'}{r}\bigg)\beta_2-\mathcal{R}\bigg(\frac{\lambda'^2}{2}+\frac{\mathcal{R}e^\xi}{2}+\frac{\lambda'}{r}\\\label{g14e}
&+\frac{\lambda'\xi'}{4}+\frac{2\xi'}{r}-\frac{\lambda''}{2}\bigg)-\mathcal{R}'\bigg(\frac{\lambda'}{4}-\frac{1}{r}
+\frac{\xi'}{4}\bigg)+\frac{\mathcal{R''}}{2}+\beta_5\bigg(\frac{\xi'}{2}+\frac{\lambda'}{2}\bigg)-\beta_8\bigg\}\bigg],\\\nonumber
P_r&=\bigg[\zeta\bigg\{\frac{3}{4}\bigg(\frac{\xi'}{r}+\frac{\lambda'}{r}\bigg)\bigg(\beta_1+\frac{\beta_2}{3}\bigg)+\frac{3}{8}\big(\xi'
+\lambda'\big)\bigg(\beta_4+\frac{\beta_5}{3}\bigg)+\mathcal{R}\bigg(\frac{5\lambda'}{4r}-\frac{\mathcal{R}e^\xi}{2}\\\nonumber
&-\frac{7\lambda'\xi'}{16}+\frac{\lambda'^2}{16}+\frac{\lambda''}{2}+\frac{\xi'}{4r}\bigg)+\mathcal{R}'\bigg(\frac{\lambda'}{8}-\frac{1}{2r}
+\frac{\xi'}{8}\bigg)-\frac{\mathcal{R''}}{4}-\frac{\beta_8}{4}\bigg\}-8\pi
e^{\xi}\bigg]^{-1}\\\nonumber
&\times\bigg[-\frac{1}{4}\bigg(\frac{\xi'}{r}+\frac{\lambda'}{r}\bigg)+8\pi\mathfrak{B_c}e^\xi-\zeta\mathfrak{B_c}\bigg\{\bigg(\frac{\xi'}{r}
+\frac{\lambda'}{r}\bigg)\beta_1+\mathcal{R}\bigg(\frac{\lambda'^2}{4}-\frac{\mathcal{R}e^\xi}{2}+\frac{\xi'}{r}\\\label{g14f}
&-\frac{\lambda'\xi'}{2}+\frac{\lambda''}{2}+\frac{2\lambda'}{r}\bigg)+\mathcal{R}'\bigg(\frac{\lambda'}{4}-\frac{1}{r}
+\frac{\xi'}{4}\bigg)-\frac{\mathcal{R''}}{2}+\beta_4\bigg(\frac{\xi'}{2}+\frac{\lambda'}{2}\bigg)-\beta_7\bigg\}\bigg].
\end{align}
We can obtain the tangential pressure corresponding to model-I by
putting energy density \eqref{g14b} and radial pressure \eqref{g14c}
in Eq.\eqref{g8b}. Similarly, use of Eqs.\eqref{g8e}, \eqref{g14e}
and \eqref{g14f} yields tangential pressure for the model-II. A
detailed study of celestial structures whose interiors are filled
with quark matter has been carried out through
$\mathbb{E}o\mathbb{S}$ \eqref{g14a} in different modified theories.
Here, we establish solutions to the field equations with respect to
both models by using this $\mathbb{E}o\mathbb{S}$, where a coupling
parameter is taken as $\zeta=\pm5$.

\section{Tolman IV Spacetime and Boundary Conditions}

To continue our analysis, we take Tolman IV spacetime in
$f(\mathcal{R},\mathcal{T},\mathcal{Q})$ scenario which gained
significant interest in the field of astronomy. This provides the
metric coefficients as
\begin{equation}\label{g15}
e^\lambda=B\left(1+\frac{r^2}{A}\right), \quad\quad
e^\xi=\frac{1+\frac{2r^2}{A}}{\left(1-\frac{r^2}{C}\right)\left(1+\frac{r^2}{A}\right)},
\end{equation}
involving a triplet ($A,B,C$) as integration constants. These
unknowns can be determined by smoothly matching the interior
geometry with exterior vacuum spacetime. Now, we check the criteria
for the acceptance of metric potentials \cite{41j}, thus their
derivatives up to second order are
\begin{align}\nonumber
\lambda'(r)&=\frac{2r}{A+r^2}, \quad
\lambda''(r)=\frac{2\big(A-r^2\big)}{\big(A+r^2\big)^2},\\\nonumber
\xi'(r)&=\frac{2r\big(AC+2Ar^2+A^2+2r^4\big)}{A^2C+r^2\big(3AC-A^2-3Ar^2+2Cr^2-2r^4\big)},\\\nonumber
\xi''(r)&=2\left[A^2C+r^2\big(3AC-A^2-3Ar^2+2Cr^2-2r^4\big)\right]^{-2}\bigg[A^3C^2+A^4C\\\nonumber
&+r^2\bigg(4A^3C+19A^2Cr^2+24ACr^4+7A^3r^2+A^4+20A^2r^4+6Ar^6\\\nonumber
&-6AC^2r^2+4Cr^6+4r^8-\frac{9A^2C^2}{2}\bigg)\bigg],
\end{align}
from where we observe that $\lambda'(0)=\xi'(0)=0,~\lambda''(0)>0$
and $\xi''(0)>0$ everywhere ($r=0$ is center of the star), hence
both metric potentials given in Eq.\eqref{g15} are acceptable.

The smooth matching of inner and outer regions at the hypersurface
provides certain conditions which are supposed to be extremely
powerful tool in understanding the whole structural formation of
celestial objects. It is important to match the nature of outer
geometry with the inner spacetime such as both regions must be
charged or uncharged, static or non-static, etc. Furthermore, in the
context of $f(\mathcal{R})$ gravity, the boundary conditions are
completely different from $\mathbb{GR}$ as a result of the inclusion
of higher-order curvature terms \cite{41jaa}, i.e., the Starobinsky
model given by $f(\mathcal{R})=\mathcal{R}+\zeta\mathcal{R}^2$ with
$\zeta$ being a restricted parameter. Nonetheless, the term
$\mathcal{R}$ in the current scenario represents $\mathbb{GR}$ while
$\mathcal{R}_{\varphi\vartheta}\mathcal{T}^{\varphi\vartheta}$ is
not contributing to the vacuum spacetime. Following this, the most
appropriate choice for the exterior geometry is the Schwarzschild
metric with $\hat{M}$ as the total mass. This has the form
\begin{equation}\label{g20}
ds^2=-\bigg(1-\frac{2\hat{M}}{r}\bigg)dt^2+\frac{dr^2}{\bigg(1-\frac{2\hat{M}}{r}\bigg)}
+r^2d\theta^2+r^2\sin^2\theta d\phi^2.
\end{equation}
Furthermore, the continuity of radial and temporal metric
coefficients of both geometries across the boundary triggers the
following limitations
\begin{eqnarray}\label{g21}
g_{tt}&{_{=}^{\Sigma}}&e^{\lambda(\mathcal{H})}=B\left(1+\frac{\mathcal{H}^2}{A}\right)=1-\frac{2\hat{M}}{\mathcal{H}},\\\label{g21a}
g_{rr}&{_{=}^{\Sigma}}&e^{\xi(\mathcal{H})}=\frac{1+\frac{2\mathcal{H}^2}{A}}{\left(1-\frac{\mathcal{H}^2}{C}\right)\left(1+\frac{\mathcal{H}^2}{A}\right)}
=\bigg(1-\frac{2\hat{M}}{\mathcal{H}}\bigg)^{-1},\\\label{g22}
\frac{\partial g_{tt}}{\partial
r}&{_{=}^{\Sigma}}&\lambda'(\mathcal{H})=\frac{2\mathcal{H}}{A+\mathcal{H}^2}=\frac{2\bar{M}\mathcal{H}}{\mathcal{H}^2\big(\mathcal{H}-2\hat{M}\big)},
\end{eqnarray}
whose simultaneous solution yields the following three unknowns as
\begin{eqnarray}\label{g23}
A=\frac{\mathcal{H}^2\left(\mathcal{H}-3\hat{M}\right)}{\hat{M}},
\quad B=1-\frac{3\hat{M}}{\mathcal{H}}, \quad
C=\frac{\mathcal{H}^3}{\hat{M}}.
\end{eqnarray}
An acceptable physical model must have zero radial pressure at the
spherical boundary ($r=\mathcal{H}$), thus the value of
$\mathfrak{B_c}$ for the model-I is obtained from Eq.\eqref{g14c} in
terms of mass and radius as
\begin{eqnarray}\label{g26}
\mathfrak{B_c}=-\frac{3\hat{M}(\mathcal{H}-2\hat{M})^2}{4\big(11\zeta\hat{M}^3-\hat{M}^2\big(8\pi\mathcal{H}^3+11\zeta\mathcal{H}\big)
+3\hat{M}\mathcal{H}^2\big(\zeta+4\pi\mathcal{H}^2\big)-4\pi\mathcal{H}^5\big)},
\end{eqnarray}
and Eq.\eqref{g14f} for the model-II lead to
\begin{align}\nonumber
\mathfrak{B_c}&=4\big[2\zeta\hat{M}^5\big(4\mathcal{H}^4+189\mathcal{H}^2-198\big)+\zeta\hat{M}^4\big(487-20\mathcal{H}^4
-634\mathcal{H}^2\big)\mathcal{H}+2\hat{M}^3\\\nonumber
&\times\big(4\pi\mathcal{H}^6+\zeta\big(8\mathcal{H}^4+184\mathcal{H}^2-91\big)\mathcal{H}^2\big)-\hat{M}^2\big(20\pi\mathcal{H}^7
+\zeta\mathcal{H}^3\big(4\mathcal{H}^4+85\mathcal{H}^2\\\label{g27}
&-19\big)\big)+2\hat{M}\big(8\pi\mathcal{H}^8+3\zeta\mathcal{H}^6\big)-4\pi\mathcal{H}^9\big]^{-1}\big[3\hat{M}\mathcal{H}^3(\hat{M}
-\mathcal{H})(\mathcal{H}-2\hat{M})^2\big].
\end{align}

The bag constant $\big(\mathfrak{B_c}\big)$ along with unknown
triplet $\big(A,~B,~C\big)$ involved in the Tolman IV spacetime can
be calculated by using the preliminary data including masses and
radii of six strange stars \cite{41k}-\cite{41m}, as can be seen in
Table $\mathbf{1}$. Table $\mathbf{2}$ presents all these constants.
It is noticed that the behavior of all these bodies is consistent
with the limit proposed by Buchdhal \cite{42a}, i.e.,
$\frac{2\hat{M}}{\mathcal{H}}<\frac{8}{9}$. As the field equations
\eqref{g8}-\eqref{g8b} and \eqref{g8c}-\eqref{g8e} (corresponding to
both models) in modified scenario are much complicated thus we apply
a constraint to construct their solution. From graphical analysis,
we find the values of the matter sector at the core as well as the
boundary, and the bag constant (through Eqs.\eqref{g26} and
\eqref{g27}) by choosing the coupling parameter as $\zeta=\pm5$, as
shown in Tables $\mathbf{3}$-$\mathbf{6}$ corresponding to models-I
and II, respectively.
\begin{table}[H]
\scriptsize \centering \caption{Masses and radii of different
compact models \cite{41k}-\cite{41m}} \label{Table1} \vspace{+0.1in}
\setlength{\tabcolsep}{0.95em}
\begin{tabular}{cccccc}
\hline\hline Star Models & $Mass(M_{\bigodot})$ & $\mathcal{H}(km)$
& $\hat{M}/\mathcal{H}$
\\\hline Cen X-3 & 1.49 & 9.51 & 0.230
\\\hline
SMC X-1 & 1.29 & 9.13 & 0.208
\\\hline
4U 1820-30 & 1.58 & 9.1 & 0.255
\\\hline
PSR J 1614 2230 & 1.97 & 10.3 & 0.281
\\\hline
PSR J 1903+327 & 1.67 & 9.82 & 0.249
\\\hline
SAX J 1808.4-3658 & 0.88 & 8.9 & 0.145  \\
\hline\hline
\end{tabular}
\end{table}
\begin{table}[H]
\scriptsize \centering \caption{Unknown triplet corresponding to
different compact models} \label{Table2} \vspace{+0.1in}
\setlength{\tabcolsep}{0.95em}
\begin{tabular}{cccccc}
\hline\hline Star Models & $A (km^{2})$ & $B$ & $C (km^{2})$
\\\hline Cen X-3 & 121.3590 & 0.3091 & 392.68
\\\hline
SMC X-1 & 151.2630 & 0.3769 & 401.33
\\\hline
4U 1820-30 & 76.0215 & 0.2343 & 324.45
\\\hline
PSR J 1614 2230 & 59.0659 & 0.1565 & 377.34
\\\hline
PSR J 1903+327 & 97.1423 & 0.2514 & 386.44
\\\hline
SAX J 1808.4-3658 & 307.3370 & 0.5639 & 544.97  \\
\hline\hline
\end{tabular}
\end{table}
\begin{table}[H]
\scriptsize \centering \caption{Values of state parameters and the
bag constant corresponding to model-I for $\zeta=5$} \label{Table3}
\vspace{+0.1in} \setlength{\tabcolsep}{0.95em}
\begin{tabular}{cccccc}
\hline\hline Star Models & $\mathfrak{B_c} (km^{-2})$ & $\mu_c
(gm/cm^3)$ & $\mu_s (gm/cm^3)$ & $P_{c} (dyne/cm^2)$
\\\hline Cen X-3 & 0.00010678 & 1.6683$\times$10$^{15}$ & 5.4945$\times$10$^{14}$ & 3.2971$\times$10$^{35}$
\\\hline
SMC X-1 & 0.00011001 & 1.3994$\times$10$^{15}$ &
5.2109$\times$10$^{14}$ & 2.4577$\times$10$^{354}$
\\\hline
4U 1820-30 & 0.00012128 & 2.5312$\times$10$^{15}$ &
5.7768$\times$10$^{14}$ & 5.6165$\times$10$^{35}$
\\\hline
PSR J 1614 2230 & 0.00009656 & 3.0690$\times$10$^{15}$ &
5.4089$\times$10$^{14}$ & 7.6149$\times$10$^{35}$
\\\hline
PSR J 1903+327 & 0.00010336 & 1.9934$\times$10$^{15}$ &
5.3527$\times$10$^{14}$ & 4.2974$\times$10$^{35}$
\\\hline
SAX J 1808.4-3658 & 0.00009107 & 7.8812$\times$10$^{14}$ & 4.4109$\times$10$^{14}$ & 9.1576$\times$10$^{34}$  \\
\hline\hline
\end{tabular}
\end{table}
\begin{table}[H]
\scriptsize \centering \caption{Values of state parameters and the
bag constant corresponding to model-II for $\zeta=5$} \label{Table4}
\vspace{+0.1in} \setlength{\tabcolsep}{0.95em}
\begin{tabular}{cccccc}
\hline\hline Star Models & $\mathfrak{B_c} (km^{-2})$ & $\mu_c
(gm/cm^3)$ & $\mu_s (gm/cm^3)$ & $P_{c} (dyne/cm^2)$
\\\hline Cen X-3 & 0.00009801 & 1.5746$\times$10$^{15}$ & 4.5861$\times$10$^{14}$ & 3.4606$\times$10$^{35}$
\\\hline
SMC X-1 & 0.00010248 & 1.3067$\times$10$^{15}$ &
4.7359$\times$10$^{14}$ & 2.6008$\times$10$^{35}$
\\\hline
4U 1820-30 & 0.00010926 & 2.3626$\times$10$^{15}$ &
4.7493$\times$10$^{14}$ & 5.9556$\times$10$^{35}$
\\\hline
PSR J 1614 2230 & 0.00008535 & 2.8657$\times$10$^{15}$ &
4.5058$\times$10$^{14}$ & 8.0189$\times$10$^{35}$
\\\hline
PSR J 1903+327 & 0.00009358 & 1.8716$\times$10$^{15}$ &
4.5861$\times$10$^{14}$ & 4.5801$\times$10$^{35}$
\\\hline
SAX J 1808.4-3658 & 0.00008789 & 7.4129$\times$10$^{14}$ & 4.1406$\times$10$^{14}$ & 9.6614$\times$10$^{34}$  \\
\hline\hline
\end{tabular}
\end{table}
\begin{table}[H]
\scriptsize \centering \caption{Values of state parameters and the
bag constant corresponding to model-I for $\zeta=-5$} \label{Table5}
\vspace{+0.1in} \setlength{\tabcolsep}{0.95em}
\begin{tabular}{cccccc}
\hline\hline Star Models & $\mathfrak{B_c} (km^{-2})$ & $\mu_c
(gm/cm^3)$ & $\mu_s (gm/cm^3)$ & $P_{c} (dyne/cm^2)$
\\\hline Cen X-3 & 0.00010597 & 1.6201$\times$10$^{15}$ & 5.4945$\times$10$^{14}$ & 3.1635$\times$10$^{35}$
\\\hline
SMC X-1 & 0.00010917 & 1.3807$\times$10$^{15}$ &
5.2109$\times$10$^{14}$ & 2.3531$\times$10$^{35}$
\\\hline
4U 1820-30 & 0.00012018 & 2.4629$\times$10$^{15}$ &
5.7768$\times$10$^{14}$ & 5.4602$\times$10$^{35}$
\\\hline
PSR J 1614 2230 & 0.00009581 & 2.9887$\times$10$^{15}$ &
5.4089$\times$10$^{14}$ & 7.3961$\times$10$^{35}$
\\\hline
PSR J 1903+327 & 0.00010256 & 1.9693$\times$10$^{15}$ &
5.3527$\times$10$^{14}$ & 4.1989$\times$10$^{35}$
\\\hline
SAX J 1808.4-3658 & 0.00009054 & 7.8812$\times$10$^{14}$ & 4.4109$\times$10$^{14}$ & 9.1576$\times$10$^{34}$  \\
\hline\hline
\end{tabular}
\end{table}
\begin{table}[H]
\scriptsize \centering \caption{Values of state parameters and the
bag constant corresponding to model-II for $\zeta=-5$}
\label{Table6} \vspace{+0.1in} \setlength{\tabcolsep}{0.95em}
\begin{tabular}{cccccc}
\hline\hline Star Models & $\mathfrak{B_c} (km^{-2})$ & $\mu_c
(gm/cm^3)$ & $\mu_s (gm/cm^3)$ & $P_{c} (dyne/cm^2)$
\\\hline Cen X-3 & 0.00012887 & 1.5746$\times$10$^{15}$ & 7.4076$\times$10$^{14}$ & 3.0205$\times$10$^{35}$
\\\hline
SMC X-1 & 0.00012779 & 1.3311$\times$10$^{15}$ &
7.2645$\times$10$^{14}$ & 2.1619$\times$10$^{35}$
\\\hline
4U 1820-30 & 0.00015371 & 2.3626$\times$10$^{15}$
&9.1361$\times$10$^{14}$ & 5.1223$\times$10$^{35}$
\\\hline
PSR J 1614 2230 & 0.00012956 & 2.8429$\times$10$^{15}$ &
7.9842$\times$10$^{14}$ & 7.0798$\times$10$^{35}$
\\\hline
PSR J 1903+327 & 0.00012938 & 1.8716$\times$10$^{15}$ &
7.8397$\times$10$^{14}$ & 4.0233$\times$10$^{35}$
\\\hline
SAX J 1808.4-3658 & 0.00009755 & 7.4129$\times$10$^{14}$ & 5.3928$\times$10$^{14}$ & 8.2449$\times$10$^{34}$  \\
\hline\hline
\end{tabular}
\end{table}

The values of the bag constant in terms of $MeV/fm^3$ analogous to
$\zeta=5$ for all compact stars are observed as
\begin{itemize}
\item Model-I:\quad $80.68,~83.13,~91.64,~72.96,~78.11$ and $68.82$
$MeV/fm^3$.
\item Model-II:\quad $74.06,~77.44,~82.56,~64.49,~70.71$ and $66.41$
$MeV/fm^3$.
\end{itemize}
and for $\zeta=-5$, we have
\begin{itemize}
\item Model-I:\quad $80.07,~82.49,~90.81,~72.39,~77.49$ and $68.42$
$MeV/fm^3$.
\item Model-II:\quad $97.38,~96.56,~116.15,~97.91,~97.76$ and $73.71$
$MeV/fm^3$.
\end{itemize}

It is evident that the above values of bag constant with respect to
all stars are found to be in its predicted range for which the
celestial bodies show stable behavior, except for the candidate 4U
1820-30. Nonetheless, several investigations have been done by
$\mathrm{CERN-SPS}$ and $\mathrm{RHIC}$, from which it is concluded
that this constant may have its value in broad range for the density
dependent bag model.

\section{Graphical Analysis of the Considered Structures}

In this sector, we inspect the diverse physical features of six
different compact bodies which are associated with fluid anisotropy
in their interiors in
$f(\mathcal{R},\mathcal{T},\mathcal{R}_{\varphi\vartheta}\mathcal{T}^{\varphi\vartheta})$
scenario. We develop solutions to the field equations
\big(\eqref{g8}-\eqref{g8b} and \eqref{g8c}-\eqref{g8e}\big) for two
modified models and present their graphical nature by taking into
account the preliminary data given in Tables $\mathbf{1}$ and
$\mathbf{2}$ corresponding to all candidates. Moreover, various
properties of massive structures such as metric potentials, energy
conditions, anisotropic pressure, mass, redshift and compactness
inside all considered candidates have been plotted to ensure whether
the resulting solutions are physically acceptable or not. We also
study the stability corresponding to both obtained models for the
considered value of $\zeta$. It is mentioned here that the
consistent solution must involve metric functions with increasing
and singularity-free nature. We observe the consistency of our
solutions from the graphs of both temporal as well as radial metric
coefficients \eqref{g15}, as shown in Figure $\mathbf{1}$.
\begin{figure}\center
\begin{tabular}{ccc}
\includegraphics[width=0.45\textwidth]{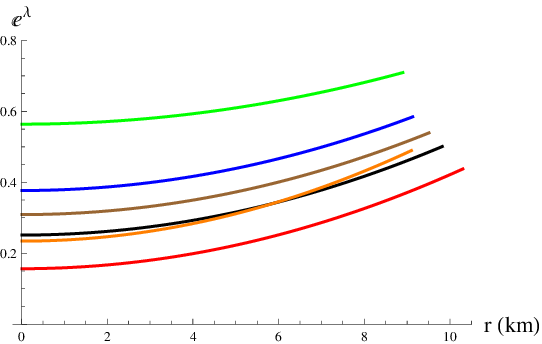} & \includegraphics[width=0.45\textwidth]{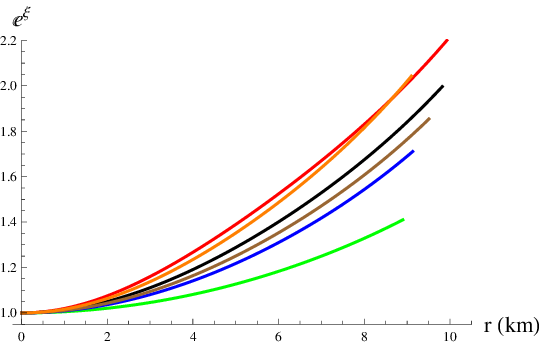}\\
\textbf{(a)} & \textbf{(b)}
\end{tabular}
\caption{Temporal \textbf{(a)} and radial \textbf{(b)} metric
coefficients versus $r$ for 4U 1820-30 (orange), Cen X-3 (brown),
SMC X-1 (blue), PSR J 1614 2230 (Red), PSR J 1903+327 (black) and
SAX J 1808.4-3658 (green).}
\end{figure}
\begin{figure}\center
\begin{tabular}{ccc}
\includegraphics[width=0.45\textwidth]{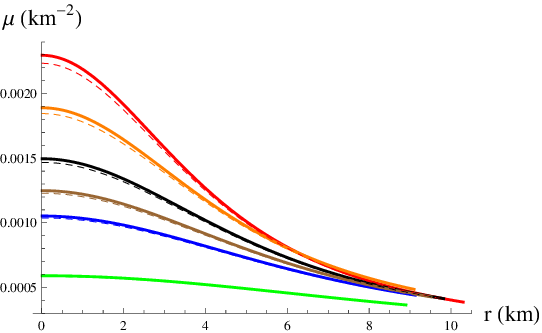} & \includegraphics[width=0.45\textwidth]{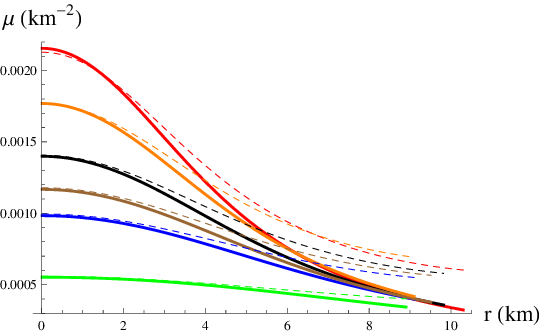}\\
\textbf{(a)} & \textbf{(b)}
\end{tabular}
\begin{tabular}{ccc}
\includegraphics[width=0.45\textwidth]{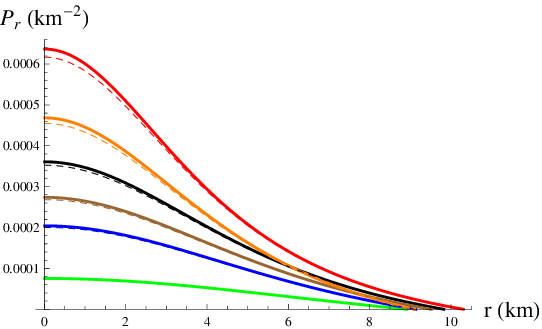} & \includegraphics[width=0.45\textwidth]{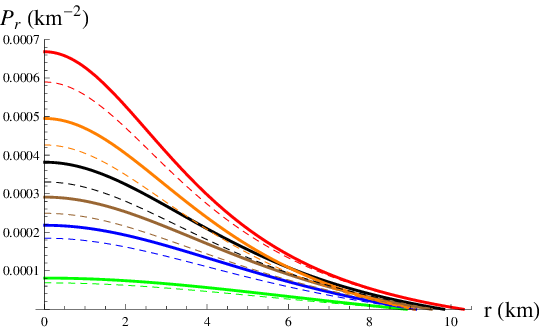}\\
\textbf{(c)} & \textbf{(d)}
\end{tabular}
\begin{tabular}{ccc}
\includegraphics[width=0.45\textwidth]{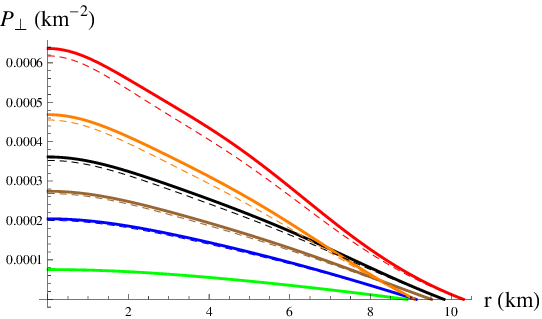} & \includegraphics[width=0.45\textwidth]{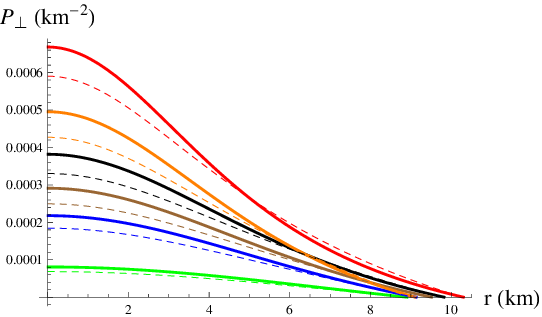}\\
\textbf{(e)} & \textbf{(f)}
\end{tabular}
\caption{Plots of the energy density \textbf{(a,b)}, radial
\textbf{(c,d)} and tangential \textbf{(e,f)} pressures versus $r$
corresponding to model-I (left) and model-II (right) for 4U 1820-30
(orange), Cen X-3 (brown), SMC X-1 (blue), PSR J 1614 2230 (Red),
PSR J 1903+327 (black) and SAX J 1808.4-3658 (green) [$\zeta=5$
(solid) and $-5$ (dashed)].}
\end{figure}

\subsection{Study of Matter Variables}

The matter inside any geometry concentrates in its core, thus the
corresponding solution will be considered physically acceptable only
if the matter variables exhibit maximum values in the middle and
monotonically decreases towards the boundary. Since the considered
fluid involves anisotropy, so we plot the effective energy density,
effective radial and tangential pressures with respect to both
models shown in Figure $\mathbf{2}$. We notice that these variables
represent acceptable behavior and hence confirms the existence of
extremely dense cores in this modified gravity. Figure $\mathbf{2}$
(first row) shows that the model-I produces more compact structures
as compared to model-II as the effective energy density takes higher
values in the former case. The other four graphs of Figure
$\mathbf{2}$ also demonstrate that radial and tangential pressures
get higher values inside each star for the model-II. Further, the
regular conditions for both models are also confirmed. However, we
do not provide their plots.

\subsection{Pressure Anisotropy}

The first solution induces anisotropic factor in the system (i.e.,
$\Delta=P_\bot-P_r$) as
\begin{align}\nonumber
\Delta&=\bigg\{\big(-\zeta  \xi ' \big(r \lambda'+4\big)+2 \zeta r
\lambda''+\zeta  r \lambda'^2+4 \zeta \lambda'+32 \pi r e^{\xi}\big)
\big(2\big(8 \zeta -\zeta r^2 \xi ''+9\zeta \\\nonumber &\times r^2
\lambda''+64 \pi  r^2 e^{\xi}-8 \zeta e^{\xi}\big)+\zeta
\big(-r^2\big) \xi '^2+r^23 \zeta  \lambda '^2-10 \zeta  r \xi '
\big(r \lambda'+4\big)+16 \zeta \\\nonumber &\times r
\lambda'\big)\bigg\}^{-1}\bigg\{3\zeta
r^3(6\zeta\mathfrak{B_c}+1)\lambda'^4+4r\big(\zeta
\xi''\big(8\mathfrak{B_c}\big(\zeta+\big(8\pi r^2-\zeta
\big)e^{\xi}\big)+r^2 (4 \zeta  \mathfrak{B_c} \\\nonumber &-1)
\lambda''\big)+8 \lambda'' \big(\zeta  (1+\zeta \mathfrak{B_c}
)+e^{\xi} \big(8 \pi  r^2 (3\zeta \mathfrak{B_c} +1)-\zeta (\zeta
\mathfrak{B_c} +1)\big)\big)+128 \pi \mathfrak{B_c}
\\\nonumber &\times e^{\xi} \big(\big(8 \pi r^2-\zeta \big)e^{\xi}+\zeta\big)+\zeta r^2 (10 \zeta \mathfrak{B_c} +9)
\lambda ''^2\big)+2 r \lambda'^2 \big(4 \big(2 \big(2 \zeta (3 \zeta
\mathfrak{B_c} +1)\\\nonumber &+(2\mathfrak{B_c} \zeta +1) \big(8
\pi r^2-\zeta \big) e^{\xi}\big)+\zeta r^2 (7 \zeta \mathfrak{B_c}
+3) \lambda''\big)-\zeta  r^2 (6 \zeta \mathfrak{B_c} +1)
\xi''\big)+4\lambda'
\\\nonumber &\times \big(8 \big(\zeta (2 \zeta
\mathfrak{B_c} +1)+e^{\xi} \big(4 \pi r^2 (8 \zeta \mathfrak{B_c}
+1)-\zeta  (2 \zeta \mathfrak{B_c}+1)\big)\big)+ r^2 (24
\zeta\mathfrak{B_c} +19)
\\\nonumber &\times\zeta\lambda''-3 \zeta r^2 \xi
''\big)+\zeta r^2 \xi '^3 \big(88 \zeta \mathfrak{B_c} +(22 \zeta
\mathfrak{B_c} r+r) \lambda'+14\big)+\zeta r \xi '^2 \big(-2
\big(8\\\nonumber &\times \big(-9 \zeta \mathfrak{B_c}
+\mathfrak{B_c} \big(\zeta +24 \pi r^2\big)e^{\xi}-4\big)+ (22 \zeta
\mathfrak{B_c} +1) \lambda ''(r)r^2\big)+9 r^2\lambda'^2 (2 \zeta
\mathfrak{B_c}\\\nonumber & +1) +6 r (8 \zeta \mathfrak{B_c} +11)
\lambda'\big)+2\zeta r^2 (16 \zeta \mathfrak{B_c} +19)
\lambda'^3-\xi ' \big(\zeta r^3 (34 \zeta \mathfrak{B_c} +13)
\lambda'^3\\\nonumber &+4 \big(\zeta  r^2 (12 \zeta \mathfrak{B_c}
+1) \xi ''+8 \big(\zeta  (2 \zeta \mathfrak{B_c} +1)+e^{\xi} \big(12
\pi r^2 (4 \zeta \mathfrak{B_c} +1)-\zeta  (1+2 \zeta\\\nonumber
&\times\mathfrak{B_c})\big)\big)+\zeta  r^2 (44 \zeta \mathfrak{B_c}
+27) \lambda''\big)+2 r \lambda' \big(\zeta r^2 (6 \zeta
\mathfrak{B_c} -1) \xi ''+8 \big(\zeta (13 \zeta \mathfrak{B_c}
+10)\\\nonumber &+e^{\xi} \big(8 \pi  r^2 (3 \zeta \mathfrak{B_c}
+1)-\zeta (\zeta \mathfrak{B_c} +1)\big)\big)+\zeta r^2 (22 \zeta
\mathfrak{B_c} +19) \lambda''\big)+2 \zeta  r^2 (92 \zeta
\mathfrak{B_c}\\\label{g31} &+35) \lambda'^2\big)\bigg\},
\end{align}
while the second solution yields
\begin{align}\nonumber
\Delta&=\bigg\{2 e^{\xi} r^2 \zeta \mathcal{R}^2+4 \zeta  \big(r
\xi'+2 e^{\xi}-r \lambda'-2\big) \mathcal{R}+r \big(-r \zeta \beta_3
\lambda'^2+\zeta \big(r \mathcal{R}'-2 r \beta_6\\\nonumber &-4
\beta_3\big) \lambda'+32 e^{\xi} \pi  r-4 r \zeta \beta_9-4 \zeta
\beta_6+\zeta \xi' \big(r \lambda' \beta_3+4 \beta_3+2 r \beta_6-r
\mathcal{R}'\big)+2 r\\\nonumber &\times \zeta \mathcal{R}''-2 r
\zeta \beta_3 \lambda''\big)\bigg\}^{-1}\bigg\{r \big(r \lambda'^2+2
\lambda'-\xi' \big(r \lambda'+2\big)+2 r
\lambda''\big)\bigg\}+\bigg\{8 e^{\xi} r \zeta
\mathcal{R}^2\\\nonumber &+\zeta \big(-r \lambda'^2-20 \lambda'+\xi'
\big(7 r \lambda'-4\big)-8 r \lambda''\big) \mathcal{R}+2 \big(2
\zeta \beta_8 r-3 \zeta \beta_4 \lambda' r-\zeta \beta_5
\lambda'\\\nonumber &\times r+2 \zeta  \mathcal{R}'' r+64 e^{\xi}
\pi r-\zeta (6 \beta_1+2 \beta_2+3 r \beta_4+r \beta_5) \xi'-\zeta
\mathcal{R}' \big(r \xi'+r \lambda'-4\big)\\\nonumber &-6 \zeta
\beta_1 \lambda'-2 \zeta \beta_2 \lambda'\big)\bigg\}^{-1}\bigg\{4
\big(2 e^{\xi} r \zeta \mathfrak{B_c} \mathcal{R}^2+\zeta
\mathfrak{B_c} \big(-r \lambda'^2-8 \lambda'+2 \xi' \big(r
\lambda'-2\big)\\\nonumber &-2 r \lambda''\big) \mathcal{R}+32
e^{\xi} \pi r \mathfrak{B_c} +4 r \zeta \mathfrak{B_c} \beta_7-4
\zeta \mathfrak{B_c}  \beta_1 \xi'-2 r \zeta \mathfrak{B_c}\beta_4
\xi'-\xi'-4 \zeta \mathfrak{B_c} \beta_1\\\nonumber &\times
\lambda'-2 r \zeta \mathfrak{B_c}  \beta_4 \lambda'-\lambda'-\zeta
\mathfrak{B_c} R' \big(r \xi'+r \lambda'-4\big)+2 r \zeta
\mathfrak{B_c} \mathcal{R}''\big)\bigg\}+\bigg\{\big(8 e^{\xi} r
\zeta \mathcal{R}^2\\\nonumber &+\zeta \big(-r \lambda'^2-20
\lambda'+\xi' \big(7 r \lambda'-4\big)-8 r \lambda''\big)
\mathcal{R}+2 \big(2 \zeta \beta_8 r-3 \zeta  \beta_4 \lambda'
r-\zeta \beta_5 \lambda' r\\\nonumber &+2 \zeta \mathcal{R}'' r+64
e^{\xi} \pi r-\zeta  (6 \beta_1+2\beta_2+3 r \beta_4+r \beta_5)
\xi'-6 \zeta \beta_1 \lambda'-2 \zeta  \beta_2 \lambda'-\zeta
\mathcal{R}'\\\nonumber &\times \big(r \xi'+r
\lambda'-4\big)\big)\big) \big(2 e^{\xi} r^2 \zeta \mathcal{R}^2+4
\zeta \big(r \xi'+2 e^{\xi}-r \lambda'-2\big) \mathcal{R}+r \big(-r
\zeta \beta_3 \lambda'^2\\\nonumber &+\zeta \big(-4 \beta_3-2 r
\beta_6+r \mathcal{R}'\big) \lambda'+32 e^{\xi} \pi r-4 r \zeta
\beta_9-4 \zeta\beta_6+\zeta \xi' \big(r \lambda' \beta_3+2 r
\beta_6\\\nonumber &+4 \beta_3-r \mathcal{R}'\big)+2 r \zeta
\mathcal{R}''-2 r \zeta \beta_3
\lambda''\big)\big)\bigg\}^{-1}\bigg\{2 r \zeta \big(\big(-r (2
\beta_1+\mathcal{R}) \lambda'^2-\big(4 \beta_1+r\\\nonumber &\times
4 \beta_4-2 r \mathcal{R}'\big) \lambda'-8 r \beta_7-8 \beta_4+\xi'
\big(2 r \lambda' \beta_1+4 \beta_1+4 r \beta_4+\mathcal{R} \big(r
\lambda'+4\big)\big)-r\\\nonumber &\times 4 \beta_1 \lambda''-2 r
\mathcal{R} \lambda''\big) \big(2 e^{\xi} r \zeta \mathfrak{B_c}
\mathcal{R}^2+\zeta \mathfrak{B_c} \big(2 r \lambda'^2+4
\lambda'+\xi' \big(r \lambda'+8\big)-2 r \lambda''\big)
\mathcal{R}\\\nonumber &-32 e^{\xi} \pi r \mathfrak{B_c} +4 r \zeta
\mathfrak{B_c}\beta_8-4 \zeta \mathfrak{B_c}  \beta_2 \xi'-2 r \zeta
\mathfrak{B_c} \beta_5 \xi'-3 \xi'-4 \zeta \mathfrak{B_c} \beta_2
\lambda'-2 r \zeta \beta_5\\\nonumber &\times \mathfrak{B_c}
\lambda'-3 \lambda'+\zeta \mathfrak{B_c} \mathcal{R}' \big(r \xi'+r
\lambda'-4\big)-2 r \zeta \mathfrak{B_c} \mathcal{R}''\big)+\big(2 r
\beta_2 \lambda'^2-r \mathcal{R} \lambda'^2+4\\\nonumber &\times
\beta_2 \lambda'+4 r \beta_5 \lambda'-4 \mathcal{R} \lambda'+8 r
\beta_8+8 \beta_5+2 \mathcal{R}' \big(r \xi'-2 r
\lambda'-4\big)+\xi' \big( (\mathcal{R}-2 \beta_2)\\\nonumber
&\times r\lambda'-4 (\beta_2+r \beta_5)\big)-4 r \mathcal{R}''+4 r
\beta_2 \lambda''-2 r \mathcal{R} \lambda''\big) \big(-2 e^{\xi} r
\zeta \mathfrak{B_c} \mathcal{R}^2+\zeta \mathfrak{B_c} \big(
\lambda'^2\\\nonumber &\times r+8 \lambda'+\xi' \big(4-2 r
\lambda'\big)+2 r \lambda''\big) \mathcal{R}-32 e^{\xi} \pi  r
\mathfrak{B_c} -4 r \zeta \mathfrak{B_c} \beta_7+4 \zeta
\mathfrak{B_c}  \beta_1 \xi'+2\\\nonumber &\times r
\zeta\mathfrak{B_c} \beta_4 \xi'+\xi'+4 \zeta \mathfrak{B_c} \beta_1
\lambda'+2 r \zeta \mathfrak{B_c}  \beta_4 \lambda'+\lambda'+\zeta
\mathfrak{B_c} \mathcal{R}' \big(r \xi'+r
\lambda'-4\big)-2\\\nonumber &\times r \zeta \mathfrak{B_c}
\mathcal{R}''\big)\big)\bigg\}.
\end{align}
We study the impact of anisotropy in the structural evolution of
considered bodies by graphically observing this factor. The nature
of anisotropy is given as follows.
\begin{itemize}
\item It shows increasing (outward) behavior when radial pressure is
less than the tangential pressure.
\item It shows decreasing (inward) behavior when tangential pressure is
less than the radial pressure.
\end{itemize}
Figure $\mathbf{3}$ exhibits that this factor vanishes at the core,
increasing outward and then again decreasing towards boundary of
each star. We can see that candidate SAX J 1808.4-3658 involves
lowest anisotropy in its interior in comparison with all other
stars.
\begin{figure}\center
\begin{tabular}{ccc}
\includegraphics[width=0.45\textwidth]{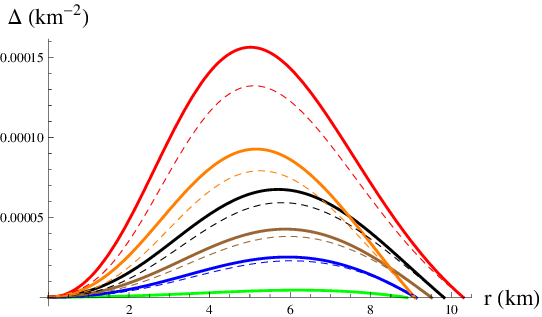} & \includegraphics[width=0.45\textwidth]{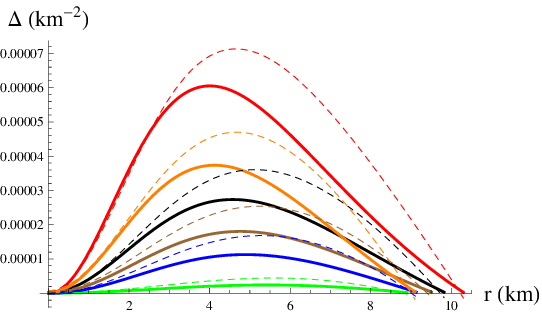}\\
\textbf{(a)} & \textbf{(b)}
\end{tabular}
\begin{tabular}{ccc}
\includegraphics[width=0.45\textwidth]{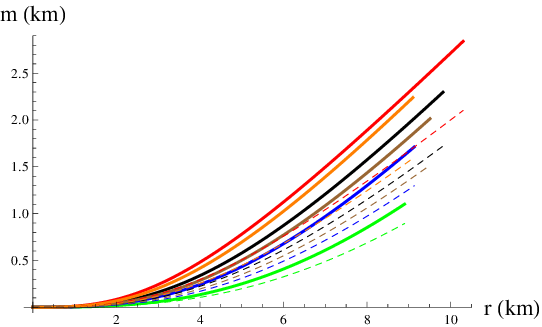} & \includegraphics[width=0.45\textwidth]{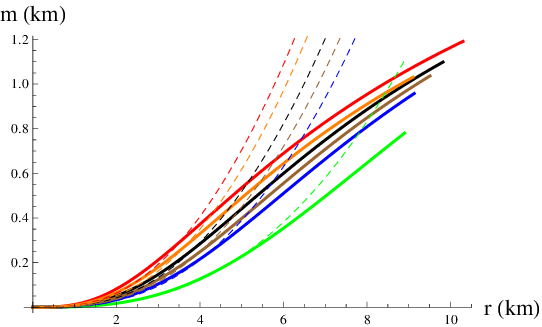}\\
\textbf{(c)} & \textbf{(d)}
\end{tabular}
\caption{Plots of anisotropy \textbf{(a,b)} and mass \textbf{(c,d)}
versus $r$ corresponding to model-I (left) and model-II (right) for
4U 1820-30 (orange), Cen X-3 (brown), SMC X-1 (blue), PSR J 1614
2230 (Red), PSR J 1903+327 (black) and SAX J 1808.4-3658 (green)
[$\zeta=5$ (solid) and $-5$ (dashed)].}
\end{figure}

\subsection{Mass, Compactness and Surface Redshift}

Another form of the mass of spherical geometry in terms of effective
energy density in this theory can be calculated as
\begin{equation}\label{g32}
m(r)=\frac{1}{2}\int_{0}^{\mathcal{H}}r^2\mu dr,
\end{equation}
where $\mu$ is presented in Eqs.\eqref{g14b} and \eqref{g14e}
corresponding to models-I and II, respectively. Alternatively, this
has the following form in accordance with Eq.\eqref{g12a} as
\begin{equation}\label{g33}
m(r)=\frac{\hat{M} r^3 \left(\hat{M} \left(r^2-3
\mathcal{H}^2\right)+2 \mathcal{H}^3\right)}{2 \mathcal{H}^3 \left(2
\hat{M} r^2-3 \hat{M} \mathcal{H}^2+\mathcal{H}^3\right)}.
\end{equation}
Figure $\mathbf{3}$ (second row) indicates that the spherical mass
exhibits increasing behavior towards boundary in the interior of
considered candidates and the solution for model-I provides more
massive compact structures.

Various physical quantities such as compactness and surface redshift
have been studied in literature to understand the evolution of
compact stars. The former is defined as the ratio between mass and
radius of a star, and is given as
\begin{align}\label{g34}
\beta(r)=\frac{m(r)}{r}=\frac{\hat{M} r^2 \left(\hat{M} \left(r^2-3
\mathcal{H}^2\right)+2 \mathcal{H}^3\right)}{2 \mathcal{H}^3 \left(2
\hat{M} r^2-3 \hat{M} \mathcal{H}^2+\mathcal{H}^3\right)}.
\end{align}
Its upper limit has been calculated which is found to be in
accordance with the proposed masses and radii of various compact
stars. Buchdahl \cite{42a} proposed its maximum value for a feasible
solution representing celestial system as $\frac{4}{9}$ after
employing the matching criteria at hypersurface ($r=\mathcal{H}$).

A heavily object surrounding with large gravitational field releases
electromagnetic radiations as a result of certain reactions
undergoing in its core. If the wavelength of those radiations
increases, one gets redshifted light. Its mathematical formula is
given by
\begin{equation}\label{g35}
z_s(r)=\frac{1}{\sqrt{1-2\beta(r)}}-1,
\end{equation}
which takes the form after inserting Eq.\eqref{g34} as
\begin{equation}\label{g36}
z_s(r)=-1+\sqrt{\frac{\mathcal{H}^3 \left(2 \hat{M} r^2-3 \hat{M}
\mathcal{H}^2+\mathcal{H}^3\right)}{\mathcal{H}^5\left(\mathcal{H}-3\hat{M}\right)+\hat{M}^2r^2\big(3\mathcal{H}^2-r^2\big)}}.
\end{equation}
The physically feasible models having isotropic inner configuration
gain its uppermost limit as $2$ which was turned to be $5.211$ for
anisotropic distribution, suggested by Ivanov \cite{42b}. Further,
the gravitational (or interior) redshift is described by
\begin{equation}\label{g351}
z_g(r)=e^{-\lambda(r)/2}-1,
\end{equation}
which must be a decreasing function of the radial coordinate towards
the boundary. Figure $\mathbf{4}$ displays the plots of all
aforementioned factors with respect to both solutions and we notice
that they lie in their required ranges. Moreover, the increment in
the bag constant increases these quantities.
\begin{figure}\center
\begin{tabular}{ccc}
\includegraphics[width=0.45\textwidth]{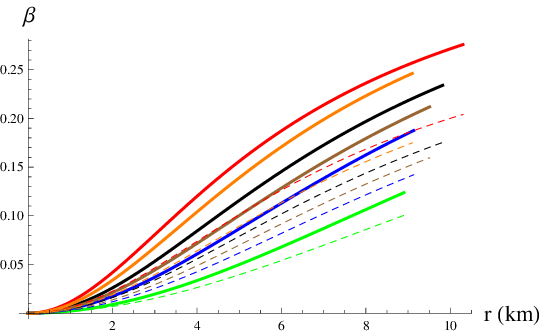} & \includegraphics[width=0.45\textwidth]{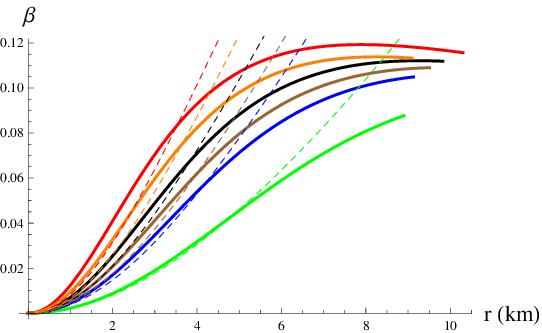}\\
\textbf{(a)} & \textbf{(b)}
\end{tabular}
\begin{tabular}{ccc}
\includegraphics[width=0.45\textwidth]{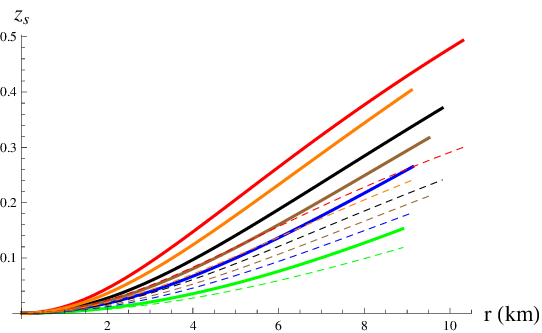} & \includegraphics[width=0.45\textwidth]{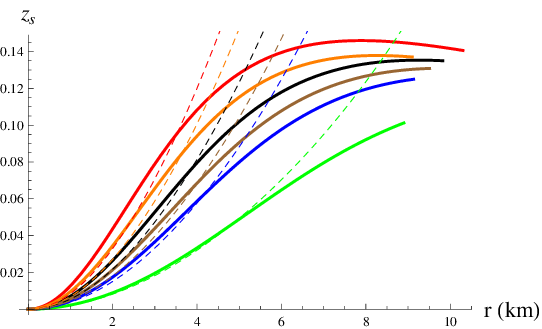}\\
\textbf{(c)} & \textbf{(d)}
\end{tabular}
\begin{tabular}{ccc}
\includegraphics[width=0.45\textwidth]{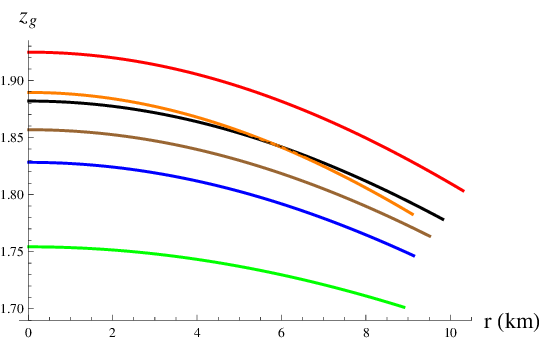}\\
\textbf{(e)}
\end{tabular}
\caption{Plots of compactness \textbf{(a,b)}, surface redshift
\textbf{(c,d)} and gravitational redshift \textbf{(e)} versus $r$
corresponding to model-I (left) and model-II (right) for 4U 1820-30
(orange), Cen X-3 (brown), SMC X-1 (blue), PSR J 1614 2230 (Red),
PSR J 1903+327 (black) and SAX J 1808.4-3658 (green) [$\zeta=5$
(solid) and $-5$ (dashed)].}
\end{figure}

\subsection{Equation of State Parameter}

Another significant factor to analyze the relevancy of the developed
self-gravitating models is the equation of state parameter. These
equations provide the relation between density and pressure of the
considered fluid distribution. Since we are working in anisotropic
scenario, they take the following form
\begin{equation}\label{g361}
\omega_r=\frac{P_r}{\mu}, \quad \omega_\bot=\frac{P_\bot}{\mu},
\end{equation}
must lie in $[0,1]$ for the effectiveness of stellar matter
configuration. Figure \textbf{5} ensures that both these parameters
are within their specified range for models-I and II.

\subsection{Energy Conditions}

The interior of a celestial structure may comprise usual or exotic
fluid. Certain conditions which depend on different physical
quantities (such as pressure and energy density) are identified as
energy conditions whose fulfilment verify the existence of normal
fluid inside a compact star. These conditions must involve the
effects of charge while studying astronomical systems influenced by
electromagnetic field. Thus, the satisfaction of the following
limitations ensure a naturalistic configuration
\begin{itemize}
\item Strong: $\mu+P_r+2P_\bot \geq 0$,
\item Weak: $\mu+P_\bot \geq 0$, \quad $\mu \geq 0$, \quad $\mu+P_r \geq 0$,
\item Null: $\mu+P_\bot \geq 0$, \quad $\mu+P_r \geq 0$,
\item Dominant: $\mu-P_\bot \geq 0$, \quad $\mu-P_r \geq 0$.
\end{itemize}
The graphical nature of all these bounds is presented in Figures
$\mathbf{6}$ and $\mathbf{7}$ in correspondence with models-I and
II, respectively. As all plots show positive behavior, thus the
developed solutions for both
$f(\mathcal{R},\mathcal{T},\mathcal{Q})$ models are physically
viable indicating normal matter in their interiors.
\begin{figure}\center
\begin{tabular}{ccc}
\includegraphics[width=0.45\textwidth]{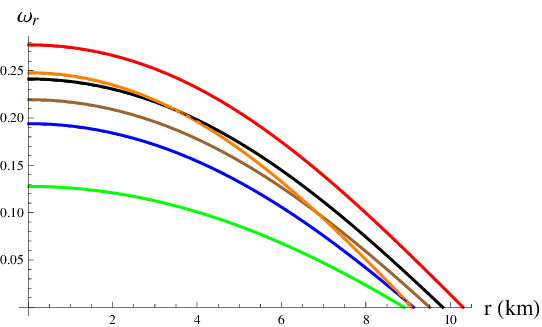} & \includegraphics[width=0.45\textwidth]{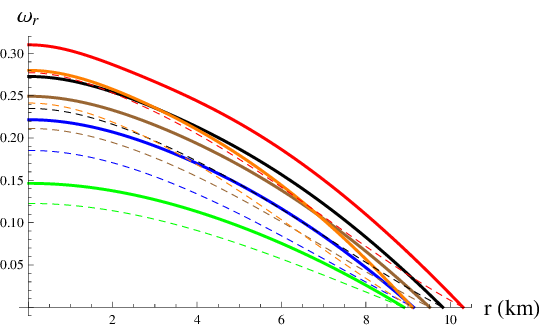}\\
\textbf{(a)} & \textbf{(b)}
\end{tabular}
\begin{tabular}{ccc}
\includegraphics[width=0.45\textwidth]{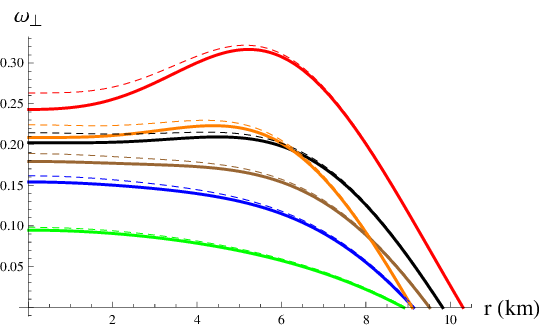} & \includegraphics[width=0.45\textwidth]{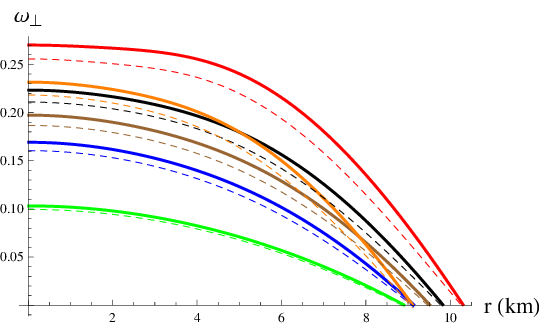}\\
\textbf{(c)} & \textbf{(d)}
\end{tabular}
\caption{Radial \textbf{(a,b)} and tangential \textbf{(c,d)}
parameters of equation of state versus $r$ corresponding to model-I
(left) and model-II (right) for 4U 1820-30 (orange), Cen X-3
(brown), SMC X-1 (blue), PSR J 1614 2230 (Red), PSR J 1903+327
(black) and SAX J 1808.4-3658 (green) [$\zeta=5$ (solid) and $-5$
(dashed)].}
\end{figure}
\begin{figure}\center
\begin{tabular}{ccc}
\includegraphics[width=0.45\textwidth]{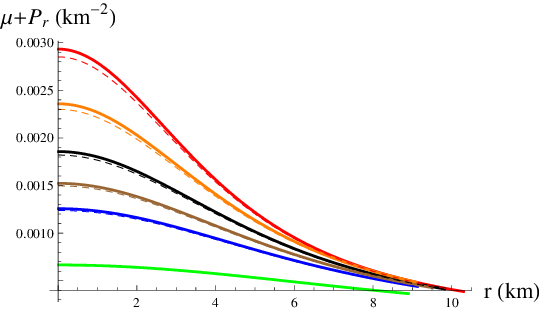} & \includegraphics[width=0.45\textwidth]{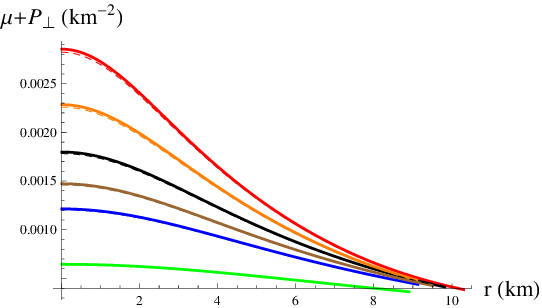}\\
\textbf{(a)} & \textbf{(b)}
\end{tabular}
\begin{tabular}{ccc}
\includegraphics[width=0.45\textwidth]{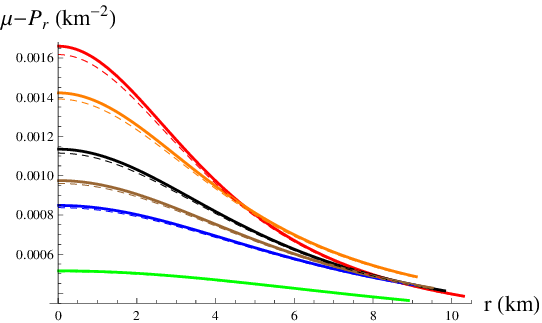} & \includegraphics[width=0.45\textwidth]{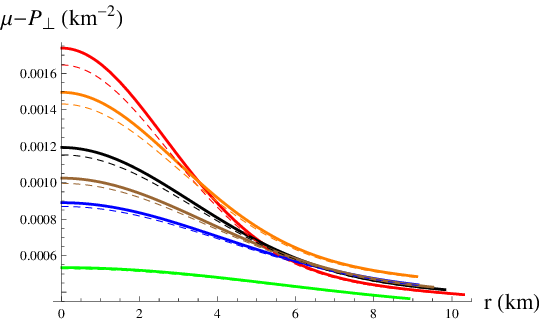}\\
\textbf{(c)} & \textbf{(d)}
\end{tabular}
\begin{tabular}{ccc}
\includegraphics[width=0.45\textwidth]{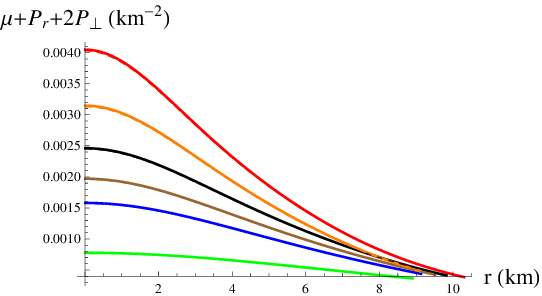}\\
\textbf{(e)}
\end{tabular}
\caption{Energy conditions \textbf{(a-e)} versus $r$ corresponding
to model-I for 4U 1820-30 (orange), Cen X-3 (brown), SMC X-1 (blue),
PSR J 1614 2230 (Red), PSR J 1903+327 (black) and SAX J 1808.4-3658
(green) [$\zeta=5$ (solid) and $-5$ (dashed)].}
\end{figure}
\begin{figure}\center
\begin{tabular}{ccc}
\includegraphics[width=0.45\textwidth]{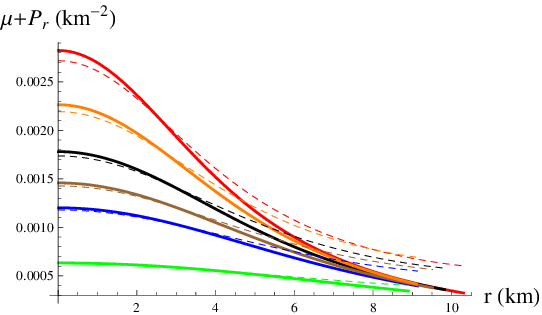} & \includegraphics[width=0.45\textwidth]{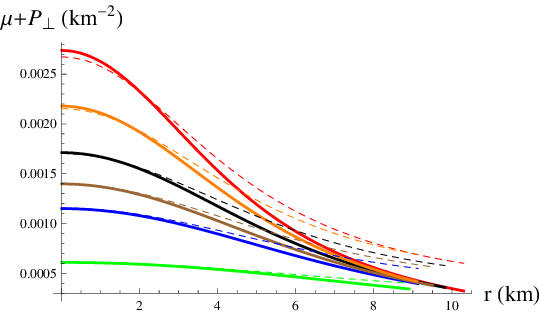}\\
\textbf{(a)} & \textbf{(b)}
\end{tabular}
\begin{tabular}{ccc}
\includegraphics[width=0.45\textwidth]{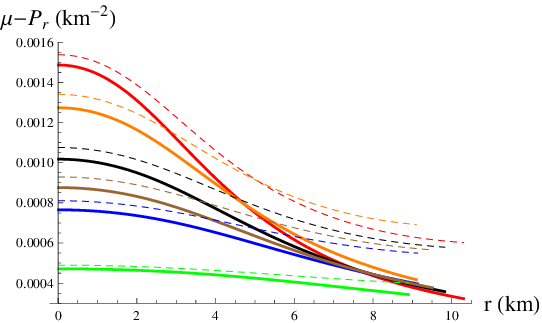} & \includegraphics[width=0.45\textwidth]{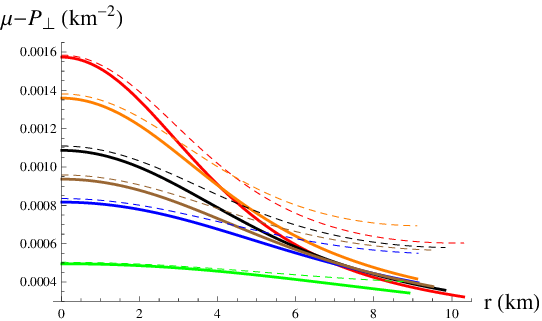}\\
\textbf{(c)} & \textbf{(d)}
\end{tabular}
\begin{tabular}{ccc}
\includegraphics[width=0.45\textwidth]{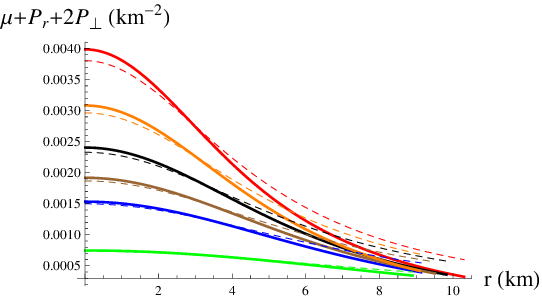}\\
\textbf{(e)}
\end{tabular}
\caption{Energy conditions \textbf{(a-e)} versus $r$ corresponding
to model-II for 4U 1820-30 (orange), Cen X-3 (brown), SMC X-1
(blue), PSR J 1614 2230 (Red), PSR J 1903+327 (black) and SAX J
1808.4-3658 (green) [$\zeta=5$ (solid) and $-5$ (dashed)].}
\end{figure}

\subsection{Tolman-Opphenheimer-Volkoff Equation}

The evolution of compact bodies can properly be figured out through
the generalized form of Tolman-Opphenheimer-Volkoff ($\mathbb{TOV}$)
equation. This subsection deduces the hydrostatic equilibrium
condition with the help of Eq.\eqref{g4a} for both models,
respectively as
\begin{align}\nonumber
&\frac{dP_r}{dr}+\frac{\lambda'}{2}\left(\mu
+P_r\right)-\frac{2}{r}\left(P_\bot-P_r\right)-\frac{2\zeta
e^{-\xi}}{\zeta\mathcal{R}+16\pi}\bigg[\frac{\lambda'\mu}{8}\bigg(\lambda'^2-\lambda'\xi'+2\lambda''+\frac{4\lambda'}{r}\bigg)\\\nonumber
&-\frac{\mu'}{8}\bigg(\lambda'^2-\lambda'\xi'+2\lambda''-\frac{4\lambda'}{r}-\frac{8e^\xi}{r^2}+\frac{8}{r^2}\bigg)
+P_r\bigg(\frac{5\lambda'^2\xi'}{8}-\frac{5\lambda'\xi'^2}{8}+\frac{7\lambda''\xi'}{4}\\\nonumber
&-\lambda'\lambda''+\frac{\lambda'\xi''}{2}-\frac{5\xi'^2}{2r}-\frac{\lambda'''}{2}+\frac{2\xi''}{r}+\frac{\lambda'\xi'}{r}-\frac{\xi'}{r^2}
-\frac{\lambda''}{r}+\frac{\lambda'}{r^2}+\frac{2e^\xi}{r^3}-\frac{2}{r^3}\bigg)-\frac{P'_r}{8}\\\nonumber
&\times\bigg(\lambda'^2-\lambda'\xi'+2\lambda''-\frac{4\xi'}{r}\bigg)+\frac{P_\bot}{r^2}\bigg(\xi'-\lambda'+\frac{2e^\xi}{r}
-\frac{2}{r}\bigg)-\frac{P'_\bot}{r}\bigg(\frac{\xi'}{2}-\frac{\lambda'}{2}+\frac{e^\xi}{r}\\\label{g11}
&-\frac{1}{r}\bigg)\bigg]=0,
\end{align}
and
\begin{align}\nonumber
&\frac{dP_r}{dr}+\frac{\lambda'}{2}\left(\mu
+P_r\right)-\frac{2}{r}\left(P_\bot-P_r\right)-\frac{2\zeta}{\zeta\mathcal{R}^2+16\pi}
\bigg[\mu\bigg\{\frac{e^{-\lambda-\xi}\lambda'\mathcal{R}\mathcal{R}_{00}}{2}-e^{-2\xi}\mathcal{R}'\\\nonumber
&\times\bigg(\frac{\lambda'}{r}-\frac{e^{\xi}}{r^2}+\frac{1}{r^2}\bigg)\bigg\}-\mu'\bigg\{\frac{e^{-\lambda-\xi}\mathcal{R}\mathcal{R}_{00}}{2}
-e^{-2\xi}\mathcal{R}\bigg(\frac{\lambda'}{r}-\frac{e^{\xi}}{r^2}+\frac{1}{r^2}\bigg)\bigg\}+P_r\\\nonumber
&\times\bigg\{\mathcal{R}'\mathcal{R}^{11}+\mathcal{R}(\mathcal{R}^{11})'-e^{-\xi}\mathcal{R}\mathcal{R}'+\frac{e^{-2\xi}\xi'
\mathcal{R}\mathcal{R}_{11}}{2}\bigg\}+P'_r\bigg\{\mathcal{R}\mathcal{R}^{11}-\frac{e^{-2\xi}\mathcal{R}\mathcal{R}_{11}}{2}\bigg\}\\\label{g11a}
&-\mathcal{R}\mathcal{R}_{22}e^{-\xi}\bigg\{\frac{P'_{\bot}}{r^2}-\frac{2P_{\bot}}{r^3}\bigg\}\bigg]=0.
\end{align}
The quantities $\mathcal{R}_{00},~\mathcal{R}_{11}$ and
$\mathcal{R}_{22}$ are given in Appendix \textbf{A}. In the
following, we check whether the resulting modified star models are
in equilibrium position or not \cite{37ccc,37ddd}. Equations
\eqref{g11} and \eqref{g11a} can be written in compact notation as
\begin{equation}\label{g36a}
f_{total}=f_a+f_h+f_m=0,
\end{equation}
where $f_a$ and $f_h$ being anisotropic and hydrostatic forces,
respectively, provided as
\begin{align}\nonumber
f_a&=\frac{2}{r}\big(P_\bot-P_r\big),\\\nonumber
f_h&=-\frac{dP_r}{dr}.
\end{align}
Further, $f_m=f_g+f_e$ contains gravitational and extra force of
this modified theory, and contains all the remaining expressions of
the above two equations with opposite sign. Figure \textbf{8}
exhibits that all the considered stellar interiors maintain their
hydrostatic equilibrium.
\begin{figure}\center
\begin{tabular}{ccc}
\includegraphics[width=0.45\textwidth]{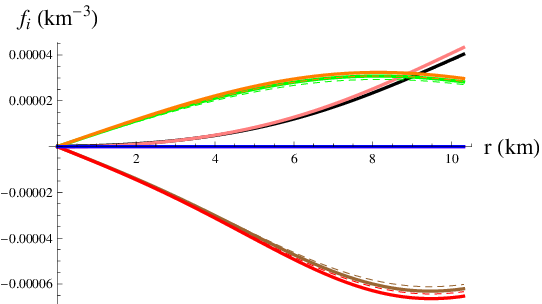} & \includegraphics[width=0.45\textwidth]{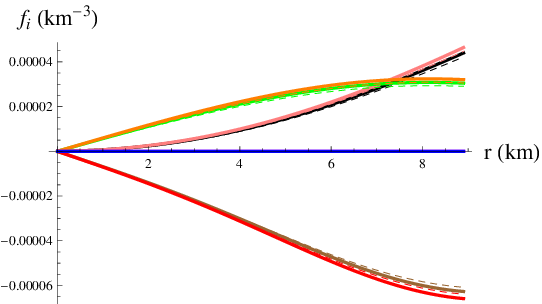}\\
\textbf{(a)} & \textbf{(b)}
\end{tabular}
\begin{tabular}{ccc}
\includegraphics[width=0.45\textwidth]{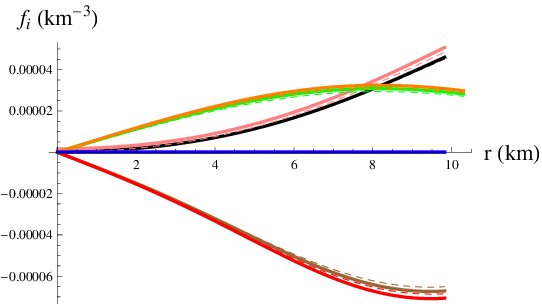} & \includegraphics[width=0.45\textwidth]{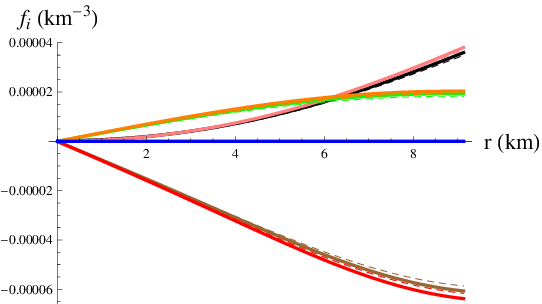}\\
\textbf{(c)} & \textbf{(d)}
\end{tabular}
\begin{tabular}{ccc}
\includegraphics[width=0.45\textwidth]{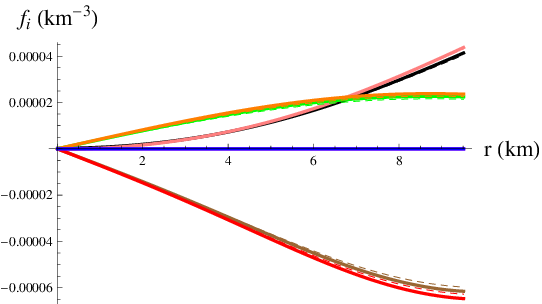} & \includegraphics[width=0.45\textwidth]{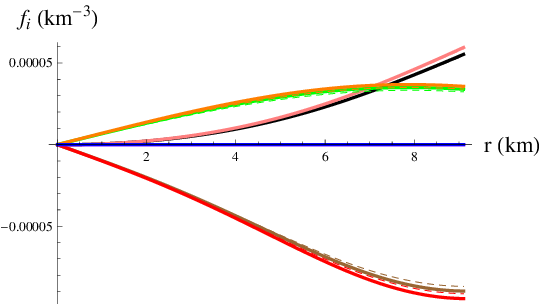}\\
\textbf{(e)} & \textbf{(f)}
\end{tabular}
\caption{Variation in $f_{m}$ (red, brown), $f_{a}$ (pink, black),
$f_{h}$ (orange, green) and $f_{total}$ (blue, magenta) for model-I
and model-II, respectively, corresponding to PSR J 1614 2230
\textbf{(a)}, SAX J 1808.4-3658 \textbf{(b)}, PSR J 1903+327
\textbf{(c)}, SMC X-1 \textbf{(d)}, Cen X-3 \textbf{(e)} and 4U
1820-30 \textbf{(f)} [$\zeta=5$ (solid) and $-5$ (dashed)].}
\end{figure}
\begin{figure}\center
\begin{tabular}{ccc}
\includegraphics[width=0.45\textwidth]{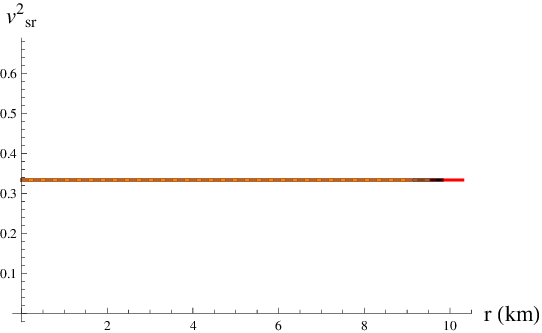} & \includegraphics[width=0.45\textwidth]{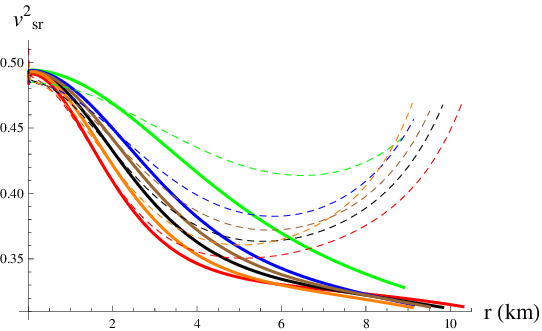}\\
\textbf{(a)} & \textbf{(b)}
\end{tabular}
\begin{tabular}{ccc}
\includegraphics[width=0.45\textwidth]{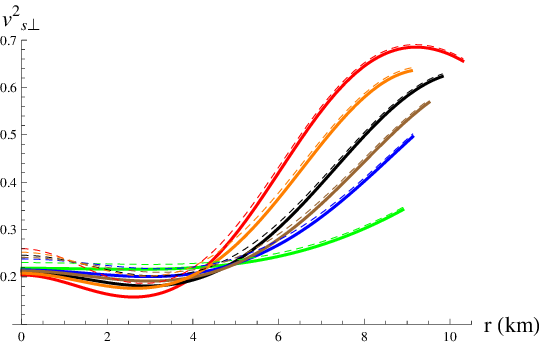} & \includegraphics[width=0.45\textwidth]{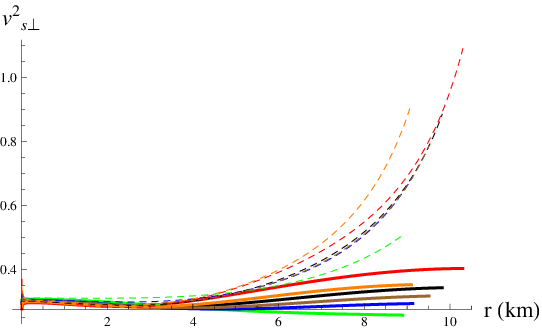}\\
\textbf{(c)} & \textbf{(d)}
\end{tabular}
\begin{tabular}{ccc}
\includegraphics[width=0.45\textwidth]{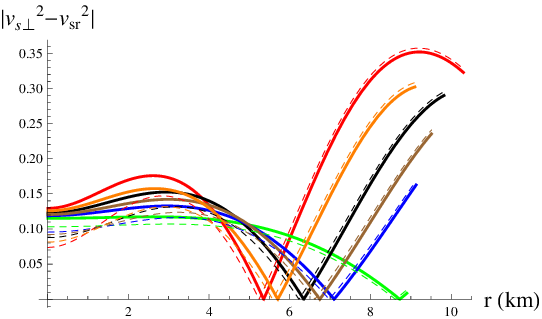} & \includegraphics[width=0.45\textwidth]{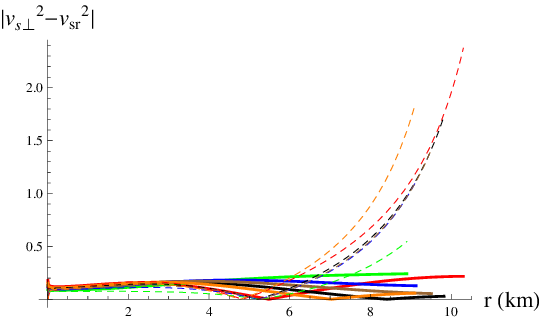}\\
\textbf{(e)} & \textbf{(f)}
\end{tabular}
\caption{Plots of radial \textbf{(a,b)} and tangential
\textbf{(c,d)} sound speeds and cracking \textbf{(e,f)} versus $r$
corresponding to model-I (left) and model-II (right) for 4U 1820-30
(orange), Cen X-3 (brown), SMC X-1 (blue), PSR J 1614 2230 (Red),
PSR J 1903+327 (black) and SAX J 1808.4-3658 (green) [$\zeta=5$
(solid) and $-5$ (dashed)].}
\end{figure}

\subsection{Stability Analysis}

\subsubsection{Causality Condition and Herrera Cracking Concept}

Amongst a large number of cosmic objects, those self-gravitating
bodies and gravitational models which satisfy stability criteria
gained considerable interest to understand their structural
composition. Multiple approaches have been mentioned in the
literature and we employ three of them to examine whether the
considered models show stable behavior or not. The speed of light in
any medium should be less than the speed of light for a stable
structure. This is known as the causality condition \cite{42d} that
can mathematically be expressed as, $0 < v_{s}^{2} < 1$. This takes
the form in anisotropic configuration as $0 < v_{s\bot}^{2} < 1$ and
$0 < v_{sr}^{2} < 1$ (sound speed in tangential and radial
directions, respectively), where
\begin{equation}
v_{s\bot}^{2}=\frac{dP_{\bot}}{d\mu}, \quad
v_{sr}^{2}=\frac{dP_{r}}{d\mu}.
\end{equation}
Herrera \cite{42e} introduced the concept of cracking and proposed
an inequality by using the above defined terms, i.e., $0 < \mid
v_{s\bot}^{2}-v_{sr}^{2} \mid < 1$ that holds only for stable
objects. Figure $\mathbf{9}$ presents graphs of the causality
condition as well as Herrera cracking approach. The fulfillment of
both conditions ensure the stability of our both models only for
$\zeta=5$. However, the compact stars except SAX J 1808.4-3658 are
unstable corresponding to model-II when $\zeta=-5$.

\subsubsection{Adiabatic Index}

The adiabatic index $\big(\Gamma\big)$ is another interesting
procedure to check the stability of compact objects. This criterion
has been employed in the analysis of self-gravitating structures and
it was found that its lower limit is $\frac{4}{3}$ everywhere in the
case of stable object \cite{42f}. Here, $\Gamma$ has the form as
\begin{equation}\label{g62}
\Gamma=\frac{\mu+P_{r}}{P_{r}}
\left(\frac{dP_{r}}{d\mu}\right)=\frac{\mu+P_{r}}{P_{r}}
\left(v_{sr}^{2}\right).
\end{equation}
The plots of $\Gamma$ (in terms of effective energy density and
pressure) for both models are depicted in Figure $\mathbf{10}$ and
we observe their stability throughout.
\begin{figure}\center
\begin{tabular}{ccc}
\includegraphics[width=0.45\textwidth]{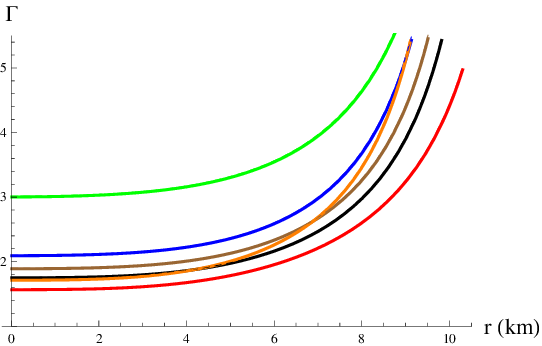} & \includegraphics[width=0.45\textwidth]{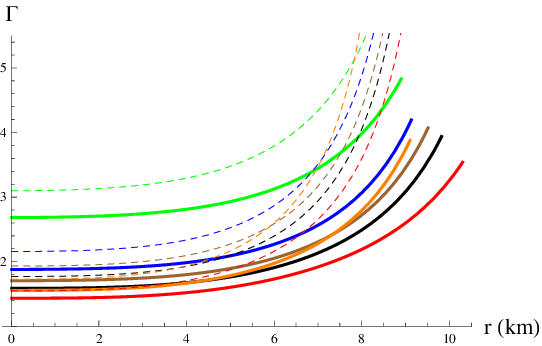}\\
\textbf{(a)} & \textbf{(b)}
\end{tabular}
\caption{Plots of adiabatic index versus $r$ corresponding to
model-I \textbf{(a)} and model-II \textbf{(b)} for 4U 1820-30
(orange), Cen X-3 (brown), SMC X-1 (blue), PSR J 1614 2230 (Red),
PSR J 1903+327 (black) and SAX J 1808.4-3658 (green) [$\zeta=5$
(solid) and $-5$ (dashed)].}
\end{figure}

\section{Conclusions}

This paper studies the existence of various compact objects which
are coupled with anisotropic distribution in the context of
$f(\mathcal{R},\mathcal{T},\mathcal{R}_{\phi\psi}\mathcal{T}^{\phi\psi})$
theory. We have adopted two standard models of this theory to
investigate the impact of non-minimal matter-geometry interaction on
the considered candidates along with $\zeta=\pm5$. The modified
field equations and the generalized $\mathbb{TOV}$ equation have
been formulated corresponding to both models. We have considered
metric potentials of the Tolman IV spacetime \eqref{g15} (that
fulfills the acceptability criteria \cite{41j}) to calculate
solutions of the systems \eqref{g8}-\eqref{g8b} and
\eqref{g8c}-\eqref{g8e}. Further, we have assumed $\mathbb{MIT}$ bag
model $\mathbb{E}$o$\mathbb{S}$ to represent the interior
distribution of strange stars. The Tolman IV spacetime involves
three unknowns ($A,B,C$) that have been calculated in terms of
masses and radii of anisotropic stars at the hypersurface. We have
analyzed six strange stars, i.e., Cen X-3,~SMC X-1,~4U 1820-30,~PSR
J 1614 2230,~PSR J 1903+327 and SAX J 1808.4-3658 whose preliminary
data (Table $\mathbf{1}$) is used to compute the unknown triplet
(Table $\mathbf{2}$) and $\mathfrak{B_c}$ with respect to both
models for $\zeta=\pm5$. We have also presented the calculated
values of the state determinants at the core and boundary of each
star which are in agreement with their observed ranges (Tables
$\mathbf{3}$-$\mathbf{6}$). The graphical interpretation of state
variables corresponding to each star model guarantees the acceptance
of both the obtained solutions \eqref{g14b}-\eqref{g14c} and
\eqref{g14e}-\eqref{g14f} (Figure $\mathbf{2}$). The mass inside all
strange stars has shown increasing behavior outwards (Figure
$\mathbf{3}$).

We have found that the model-I produces more dense interiors as
compared to the model-II for the considered values of $\zeta$. The
ranges of compactness and redshift factors are found to be
acceptable (Figure $\mathbf{4}$). The energy bounds shows positive
behavior throughout (Figures $\mathbf{6}$ and $\mathbf{7}$), hence
both of our developed solutions are physically viable. Finally, we
have used three criteria to check whether the resulting solutions
are stable or not in this modified framework, i.e., causality
conditions, Herrera's cracking technique. The graphical nature of
both approaches provides desirable results only for $\zeta=5$, hence
the solutions \eqref{g14b}-\eqref{g14c} and
\eqref{g14e}-\eqref{g14f} are stable (Figures $\mathbf{9}$ and
$\mathbf{10}$) and consistent with \cite{27a,38}. It is found that
our solutions are more effective in comparison with \cite{25a} and
thus concluded that extra force in this theory may provide better
results for the positive parametric value. On the other hand, all
the considered stars except SAX J 1808.4-3658 are found to be
unstable corresponding to model-II for $\zeta=-5$. Our results can
ultimately be reduced to $\mathbb{GR}$ for $\zeta=0$.

\section*{Appendix A}

The values of $\beta_i^{'s}$ appearing in the field equations
\eqref{g8c}-\eqref{g8e} are
\begin{align}\nonumber
\beta_1&=\frac{e^{-\xi}}{4}\bigg(\lambda'^2-\lambda'\xi'+2\lambda''+\frac{4\lambda'}{r}\bigg),\\\nonumber
\beta_2&=\frac{e^{-\xi}}{4}\bigg(\lambda'\xi'-\lambda'^2-2\lambda''+\frac{4\xi'}{r}\bigg),\quad
\beta_3=e^{-\xi}\bigg(\frac{\xi'}{r}-\frac{\lambda'}{r}+\frac{2e^\xi}{r^2}-\frac{2}{r^2}\bigg),\\\nonumber
\beta_4&=\frac{e^{-\xi}}{4}\bigg(\lambda'\xi'^2-\lambda'^2\xi'-3\lambda''\xi'-\lambda'\xi''+2\lambda'\lambda''+2\lambda'''
-\frac{4\lambda'\xi'}{r}+\frac{4\lambda''}{r}-\frac{4\lambda'}{r^2}\bigg),\\\nonumber
\beta_5&=\frac{e^{-\xi}}{4}\bigg(\lambda'^2\xi'-\lambda'\xi'^2+3\lambda''\xi'+\lambda'\xi''-2\lambda'\lambda''-2\lambda'''-\frac{4\xi'^2}{r}
+\frac{4\xi''}{r}-\frac{4\xi'}{r^2}\bigg),\\\nonumber
\beta_6&=e^{-\xi}\bigg(\frac{\lambda'\xi'}{r}-\frac{4\xi'e^\xi}{r^2}-\frac{\xi'^2}{r}+\frac{\xi''}{r}+\frac{\xi'}{r^2}-\frac{\lambda''}{r}
+\frac{\lambda'}{r^2}-\frac{4e^\xi}{r^3}+\frac{4}{r^3}\bigg),\\\nonumber
\beta_7&=\frac{e^{-\xi}}{4}\bigg(\lambda'^2\xi'^2-\lambda'\xi'^3+4\lambda''\xi'^2+3\lambda'\xi'\xi''-5\lambda'''\xi'
-4\lambda''\xi''-\lambda'^2\xi''-\lambda'\xi'''\\\nonumber
&-4\lambda'\lambda''\xi'+2\lambda''^2+2\lambda'\lambda'''+2\lambda''''+\frac{4\lambda'\xi'^2}{r}
-\frac{8\lambda''\xi'}{r}-\frac{4\lambda'\xi''}{r}+\frac{4\lambda'''}{r}-\frac{8\lambda''}{r^2}\\\nonumber
&+\frac{8\lambda'}{r^3}\bigg),\\\nonumber
\beta_8&=\frac{e^{-\xi}}{4}\bigg(\lambda'\xi'^3-\lambda'^2\xi'^2-4\lambda''\xi'^2-3\lambda'\xi'\xi''+4\lambda'\lambda''\xi'+5\lambda'''\xi'
+\lambda'^2\xi''+\lambda'\xi'''\\\nonumber
&+4\lambda''\xi''-2\lambda''^2-2\lambda'\lambda'''-2\lambda''''+\frac{4\xi'^3}{r}
-\frac{12\xi'\xi''}{r}+\frac{8\xi'^2}{r^2}+\frac{4\xi'''}{r}-\frac{8\xi''}{r^2}\\\nonumber
&+\frac{8\xi'}{r^3}\bigg),\\\nonumber
\beta_9&=e^{-\xi}\bigg(\frac{\xi'^3}{r}-\frac{\lambda'\xi'^2}{r}-\frac{3\xi'\xi''}{r}+\frac{4\lambda''\xi'}{r}-\frac{2\lambda'\xi'}{r^2}
-\frac{6\xi'}{r^3}+\frac{\lambda'\xi''}{r}-\frac{\xi''e^\xi}{r^2}+\frac{8\xi'e^\xi}{r^3}\\\nonumber
&+\frac{\xi'''}{r}-\frac{\lambda'''}{r}+\frac{2\lambda''}{r^2}-\frac{2\lambda'}{r^3}+\frac{12e^\xi}{r^4}
-\frac{12}{r^4}\bigg).
\end{align}
The geometric terms (Ricci scalar and Ricci tensors) for metric
\eqref{g6} are
\begin{align}\nonumber
\mathcal{R}&=\frac{1}{2e^{\xi}}\bigg(\lambda'^2-\xi'\lambda'+2\lambda''-\frac{4\xi'}{r}+\frac{4\lambda'}{r}-\frac{4e^{\xi}}{r^2}
+\frac{4}{r^2}\bigg),\\\nonumber
\mathcal{R}_{00}&=\frac{1}{4e^{\xi-\lambda}}\bigg(\lambda'^2-\xi'\lambda'+2\lambda''+\frac{4\lambda'}{r}\bigg),\\\nonumber
\mathcal{R}_{11}&=-\frac{1}{4}\bigg(\lambda'^2-\xi'\lambda'+2\lambda''-\frac{4\xi'}{r}\bigg),
\quad
\mathcal{R}_{22}=\frac{1}{2e^{\xi}}\bigg(\xi'r-\lambda'r+2e^\xi-2\bigg).
\end{align}
\\\\
\textbf{Data availability:} No new data were generated or analyzed
in support of this research.

\end{document}